\documentclass[aps,pra,twocolumn,showpacs,superscriptaddress,shortbibliography]{revtex4-2}
\usepackage{graphicx} 
\usepackage{amsmath}
\usepackage{soul}
\usepackage{graphicx,epstopdf}
\usepackage{gensymb}
\usepackage[dvipsnames]{xcolor}
\usepackage{footnote}
\usepackage{multibib}

\newcommand{\be}{\begin{equation}}
	\newcommand{\ee}{\end{equation}}
\newcommand{\bea}{\begin{eqnarray}}
	\newcommand{\eea}{\end{eqnarray}}
\newcommand{\bse}{\begin{subequations}}
	\newcommand{\ese}{\end{subequations}}

\graphicspath{{../}}
\usepackage{color}
\usepackage[colorlinks,bookmarks=false,citecolor=darkblue,linkcolor=red,urlcolor=blue]{hyperref}

\definecolor{darkred}{rgb}{0.7,0.0,0.0}

\definecolor{darkblue}{rgb}{0,0.02,0.45}

\definecolor{darkgreen}{rgb}{0.02,0.45,0.0}

\definecolor{violet}{rgb}{0.8,0.2,0.6}

\begin{document}
\title{Cluster-glass behaviour and large magnetocaloric effect in frustrated hyperkagome ferromagnet Li$_2$MgMn$_3$O$_8$}
\author{R. Kolay}
\author{A. Magar}
\affiliation{School of Physics, Indian Institute of Science Education and Research Thiruvananthapuram-695551, India}
\author{A. A. Tsirlin}
\affiliation{Felix Bloch Institute for Solid-State Physics, Leipzig University, 04103 Leipzig, Germany}	
\author{R. Nath}
\email{rnath@iisertvm.ac.in}
\affiliation{School of Physics, Indian Institute of Science Education and Research Thiruvananthapuram-695551, India}
\date{\today}

\begin{abstract}
A detailed study of the structural and magnetic properties of the spin-$3/2$ hyperkagome lattice compound Li$_2$MgMn$_3$O$_8$ is reported. This material shows ferromagnetic response below $T_{\rm C} \simeq 20.6$~K, the temperature almost three times lower than the Curie-Weiss temperature $\theta_{\rm CW} \simeq 56.6$~K. Density-functional band-structure calculations suggest that this reduction in $T_{\rm C}$ may be caused by long-range antiferromagnetic couplings that frustrate nearest-neighbor ferromagnetic couplings on the hyperkagome lattice. Large magnetocaloric effect is observed around the $T_{\rm C}$ with a maximum value of isothermal entropy change $\Delta S_{\rm m}\simeq 20$~J/kg-K and a maximum relative cooling power of $RCP\simeq 840$~J/kg for the 7~T magnetic field change. Critical analysis of the magnetization data and scaling analysis of the magnetocaloric effect suggest the 3D Heisenberg/XY universality class of the transition. The DC and AC magnetization measurements further reveal glassy nature of the ferromagnetic transition.
A detailed study of the non-equilibrium dynamics via magnetic relaxation and memory effect measurements demonstrates that the system evolves through a large number of intermediate metastable states and manifests significant memory effect in the cluster-glass state.
\end{abstract}

\maketitle
\section{Introduction}
Magnetic frustration arising either from the underlying lattice geometry or competing magnetic interactions fosters many nontrivial ground states, e.g., quantum spin liquid (QSL), spin ice, spin glass (SG), etc~\cite{Starykh052502,Ramirez453}. 
The cubic spinels with the general formula $AB_2$O$_4$ are a well-known example of frustrated magnets in 3D, where the magnetic $B$ site forms a pyrochlore structure~\cite{Lee011004}. For example, HgCr$_2$O$_4$ demonstrates an exotic spin-liquid-like ground state~\cite{Tomiyasu035115}. Another interesting state hosted by this family of compounds is SG as observed in LiMn$_2$O$_4$ and CoAl$_2$O$_4$~\cite{Jang2504,Hanashima024702}.
SG is a disordered ground state formed by randomly frozen spins. The primary causes of the glassy behavior are dilute magnetic impurities and structural disorder, especially in combination with magnetic frustration~\cite{Mydosh280,Binder801}. Experimentally, SG systems provide a fertile ground to study the exchange bias, magnetic memory effect, and magnetic relaxation~\cite{Jonason3243,Fisher373,Nayak127204}.

The cubic spinel structure can be modified to obtain the hyperkagome geometry, another frustrated lattice in 3D. This 3D lattice formed by corner-sharing triangles gives rise to spin-liquid physics in the case of antiferromagnetic couplings~\cite{Chillal2348,Okamoto137207}. It is featured by modified spinel compounds $AM_{0.5}$Mn$_{1.5}$O$_{4}$ ($A$ = Li, Cu and $M$ = Ni, Mg, Zn) with nonmagnetic ions on the $M$ site~\cite{Branford1649}. 
Such compounds retain cubic symmetry with the space group $P4_332$. Preliminary magnetic measurements revealed ferromagnetic (FM) behavior of LiZn$_{0.5}$Mn$_{1.5}$O$_4$ and LiMg$_{0.5}$Mn$_{1.5}$O$_4$, whereas CuNi$_{0.5}$Mn$_{1.5}$O$_4$ and LiNi$_{0.5}$Mn$_{1.5}$O$_4$ are ferrimagnets~\cite{Branford1649}. Ferrimagnetism was also observed in LiNi$_{0.5}$Mn$_{1.5}$O$_4$ (LNMO) studied in the nano-crystalline form~\cite{Islam134433}.
The bulk thermodynamic, neutron diffraction, and NMR measurements confirmed the onset of ferrimagnetism below $T_{\rm C} \simeq 125$~K with the 3D XY-type critical behavior. This material reveals unconventional superparamagnetic behavior and magnetic memory effect. It further shows a large magnetocaloric effect (MCE) as well as enhanced cooling power around $T_{\rm C}$.

Spinel materials may be promising for magnetocaloric applications, especially in the sub-Kelvin temperature range for basic research, and in the intermediate temperature range for hydrogen liquefaction~\cite{Franco112} 
where they could replace the costly and rare cryogenic liquids, such as helium~\cite{Gschneidner1479}. Ferromagnetic compounds are especially interesting in this respect because they show a large change in the magnetic entropy upon a small change in the applied field. The operation temperature of such materials is determined by their Curie temperature $T_{\rm C}$. Reducing $T_{\rm C}$ without diminishing the magnetic entropy density remains a challenging problem for low-temperature applications. In antiferromagnets, magnetic frustration has been instrumental in suppressing magnetic transitions and shifting the operation range of a magnetocaloric material toward low temperatures~\cite{Zhitomirsky104421,Tokiwa42}. However, frustration is less common in ferromagnets because ferromagnetic couplings are not in competition with each other.

\begin{figure}
 \includegraphics[width=\columnwidth]{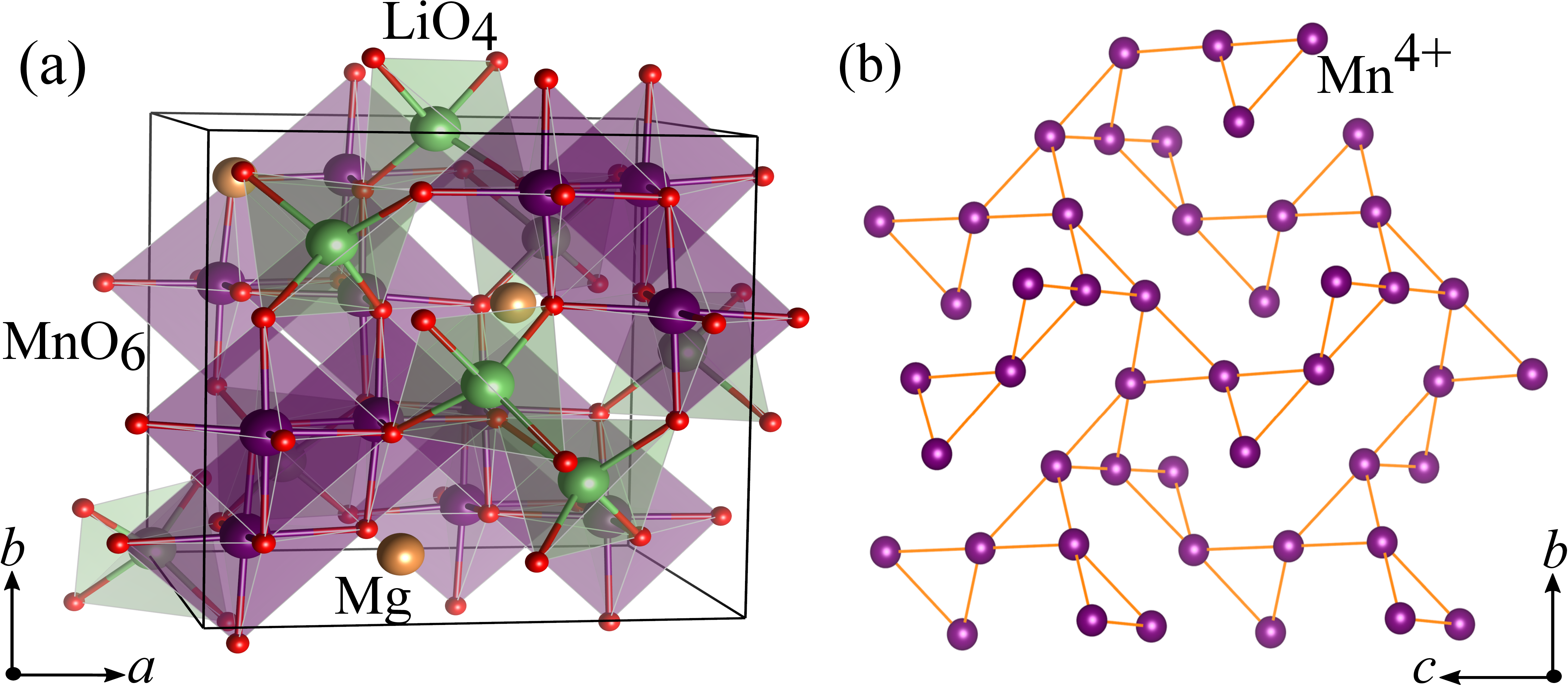}
\caption{\label{Fig1}(a) Unit cell of LMMO formed by the edge shared MnO$_6$ octahedra and LiO$_4$ tetrahedra units. (b) The 3D hyperkagome network of Mn$^{4+}$ ions.}
\end{figure}

In this paper, we report Li$_2$MgMn$_3$O$_8$ (or LiMg$_{0.5}$Mn$_{1.5}$O$_4$) (LMMO) as a ferromagnetic compound with the large entropy density and reduced $T_{\rm C}$. Unlike LNMO, LMMO features Mn$^{4+}$ as the only magnetic ion. Figure~\ref{Fig1}(a) presents the crystal structure of LMMO where MnO$_6$ octahedron are connected either by direct edge sharing or via LiO$_4$ tetrahedra. The resultant Mn$^{4+}$ hyperkagome lattice is shown in Fig.~\ref{Fig1}(b). Magnetization and heat capacity measurements confirm ferromagnetic response below $T_{\rm C} \simeq 20.6$~K. However, this value of $T_{\rm C}$ appears to be reduced as a result of frustration by interactions beyond nearest neighbors, whereas the state below $T_{\rm C}$ is better described as 
cluster glass with memory effect. Moreover, our MCE measurements put LMMO forward as an excellent candidate for magnetic refrigeration.

\section{Methods}
A polycrystalline sample of LMMO was prepared using the conventional solid-state method. Stoichiometric amounts of Li$_2$CO$_3$ (Sigma Aldrich, $\geq$ 99.9\%), MnO$_2$ (Sigma Aldrich, $\geq$ 99.9\%), and MgO (Sigma Aldrich, $\geq$ 99.99\%) were ground thoroughly for several hours and pressed into pellets. The pellets were placed in a crucible and fired at 900~\degree C for 12~h and at 700~\degree C for 48~h with intermediate grindings. X-ray diffraction (XRD) data were collected on the PANalytical powder x-ray diffractometer (Cu$K_\alpha$, $\lambda_{\rm avg}= 1.5418$ \AA) at room temperature, as well as over a wide temperature range (13~K~$\leq T \leq 300$~K) using an Oxford-Phenix low-temperature attachment.

DC magnetization ($M$) as a function of temperature was measured using a superconducting quantum interference device (SQUID) (MPMS-3, Quantum Design) magnetometer in the temperature range of 1.8~K to 380~K. Isothermal magnetization [$M(H)$] was recorded by varying the magnetic field up to 7~T at different temperatures. Similarly, AC susceptibility was measured by varying the temperature (2~K~$\leq T \leq 100$~K) and frequency (100~Hz~$ \leq \nu \leq 10$~kHz) in an applied AC field of 5~Oe using the ACMS option of the PPMS. Magnetic relaxation and magnetic memory effect were also investigated using the VSM option of the PPMS. Heat capacity ($C_{\rm p}$) as a function of temperature was measured on a small sintered pellet using the thermal relaxation technique in PPMS in the applied fields from 0 to 9~T.

Exchange couplings were evaluated for the spin Hamiltonian
\begin{equation}
 \mathcal H=\sum_{\langle ij\rangle}J_{ij}\mathbf S_i\mathbf S_j,
\end{equation}
where $S=3/2$ and the summation is over atomic pairs. The $J_{ij}$ values were obtained by a mapping procedure~\cite{xiang2011,tsirlin2014} using density-functional (DFT) band-structure calculations performed in the \texttt{VASP} code~\cite{vasp1,vasp2} with the Perdew-Burke-Ernzerhof flavor of the exchange-correlation potential~\cite{pbe96}. Correlation effects in the Mn $3d$ shell were treated within the mean-field DFT+$U$ procedure with the on-site Coulomb repulsion $U_d=2$\,eV, Hund's coupling $J_d=1$\,eV, and double-counting correction in the atomic limit. While the optimal $U_d$ value is quite low in this case, it is consistent with $U_d=2-3$\,eV used for the microscopic modeling of Cr$^{3+}$ compounds with the same $d^3$ electronic configuration of the transition-metal ion~\cite{janson2013,janson2014}. Experimental structural parameters of LMMO~\cite{Branford1649} and Zn$_2$Mn$_3$O$_8$~\cite{kitani2021} were used in the calculations. A fully ordered structure was assumed in the LMMO case.

\section{Results and Discussion}
\subsection{X-ray Diffraction}
\begin{figure}
\includegraphics[width=\columnwidth]{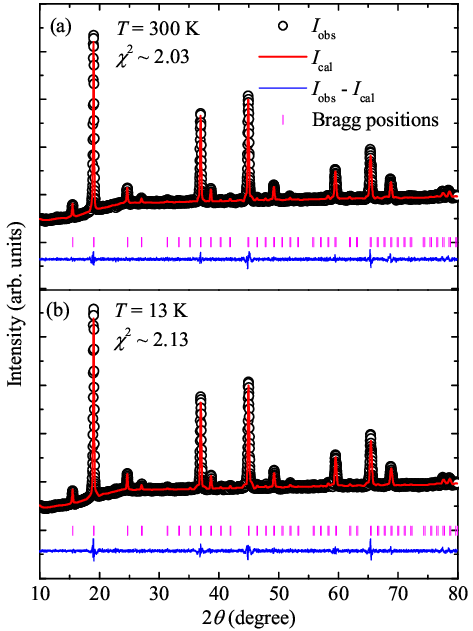}
\caption{\label{Fig2} (a) Room temperature ($T=300$~K) powder XRD pattern of the titled compound. The open circles represent the experimental data and the solid red line is the Rietveld refined fit. Expected Bragg positions are shown in pink vertical bars and the bottom line indicates the difference between observed and experimental intensities. (b) XRD pattern along with Rietveld fit at $T = 13$~K. The value of $\chi^2$ indicates the goodness of fit.}
\end{figure}
\begin{figure}
\includegraphics[width=\columnwidth]{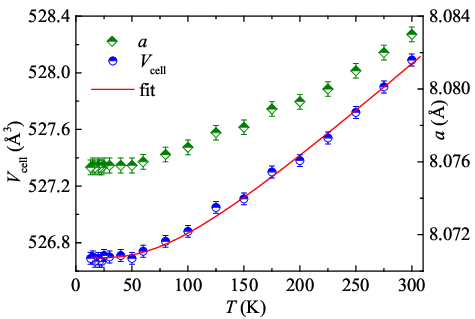}
\caption{\label{Fig3} Temperature variation of the lattice constant ($a$) and unit-cell volume ($V_{\rm cell}$). The solid red line represents the fit of $V_{\rm cell}(T)$ using Eq.~\eqref{eq1}.}
\end{figure}
Figure~\ref{Fig2}(a) displays powder XRD patterns at 300~K and 13~K, along with the Rietveld refinement fits. Both patterns can be indexed using a cubic structure with the space group $P4_{3}32$. The room-temperature lattice parameters $a = b = c = 8.083(1)$~\AA\ and unit cell volume $V_{\rm cell} \simeq 528.09$~\AA$^3$ are in a good agreement with the previous report~\cite{Branford1649}. The lattice constant of LMMO decreases monotonically on cooling (Fig.~\ref{Fig3}). $V_{\rm cell}(T)$ can be fitted by the equation~\cite{Sebastian104425,Mohanty104424}
\begin{equation}
V_{\rm cell} = \frac{\gamma\, U(T)}{K_{0}}+V_{0},
\label{eq1}
\end{equation}
where $V_{0}$ stands for the unit-cell volume at $T$ = 0~K, $K_{0}$ is the bulk modulus, and $\gamma$ is the Gr{\"u}neisen parameter. The internal energy of the system, $U(T)$, can be expressed using the Debye approximation,
\begin{equation}
U(T) = 9p\,k_{\rm B}T \left(\frac{T}{\theta_{\rm D}}\right)^3 \int_{0}^{\theta_{\rm D}/T}\frac {x^3}{(e^x-1)} \,dx. \
\end{equation}
Here, $p$ is the number of atoms in the unit cell, and $k_{\rm B}$ is the Boltzmann constant. The fitting returns the Debye temperature $\theta_{\rm D} \simeq 330$~K, $\frac{\gamma}{K_{0}} \simeq 2.02 \times 10^{-4}$~Pa$^{-1}$, and $V_{0}\simeq 526.69$~\AA$^3$.

\subsection{Magnetization}
\begin{figure}
\includegraphics[width=\columnwidth]{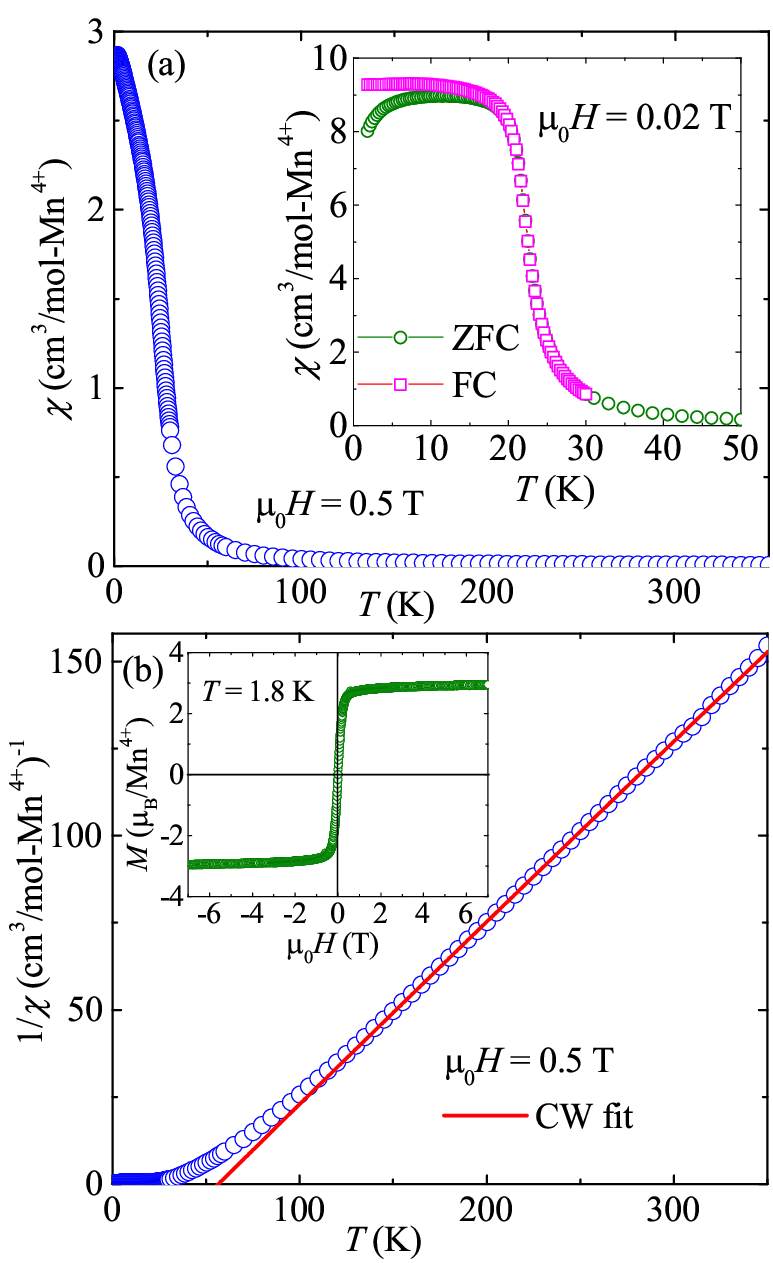}
\caption{\label{Fig4} (a) $\chi$ as a function of temperature measured in an applied magnetic field of $\mu_{0}H = 0.5$~T. Inset: $\chi(T)$ measured at $\mu_{0}H = 0.02$~T in both ZFC and FC protocols. (b) Inverse susceptibility $1/\chi$ vs $T$ along with the CW fit (solid line). Inset: A complete magnetic isotherm measured at $T=1.8$~K.}
\end{figure}
DC magnetic susceptibility $\chi~(\equiv M/H)$ as a function of temperature measured on the powder sample of LMMO in an applied field of $\mu_{0}H = 0.5$~T is depicted in Fig.~\ref{Fig4}(a). As temperature decreases, $\chi(T)$ increases slowly and shows an upturn at around 50~K, implying the onset of ferromagnetic/ferrimagnetic correlations. From $d\chi/dT$ vs $T$ plot (not shown here), the ordering temperature is found to be around $T_{\rm C} \simeq 20$~K. However, $\chi(T)$ measured under zero-field-cooled (ZFC) and field-cooled (FC) conditions [inset of Fig.~\ref{Fig4}(a)] show a bifurcation below the transition temperature in a small applied field of $\mu_{0}H = 0.02$~T. This kind of irreversibility is characteristic of systems with glassy dynamics or superparamagnetic (SP) behavior~\cite{Nath224513,Islam134433,Bag144436}. The FC $\chi(T)$ of a SG system typically remains flat or decreases with decreasing temperature below the bifurcation point, whereas in SP system it shows an increasing trend on cooling~\cite{Bandyopadhyay214410,Sasaki104405}. The LMMO data are suggestive of the SG scenario, as we confirm by the detailed study of the AC susceptibility, which is discussed later.

To extract the magnetic parameters, the high-temperature part of $1/\chi$ is fitted by the Curie-Weiss (CW) law, 
\begin{equation}
\chi(T) = \chi_{0}+\frac{C}{(T-\theta_{\rm CW})},
\end{equation}
where $\chi_{0}$ is the $T$-independent susceptibility, $C$ is the Curie constant, and $\theta_{\rm CW}$ is the characteristic Curie-Weiss temperature. The fit is shown in Fig.~\ref{Fig4}(b) for $T\geq 150$~K that returns $\chi_{0}\simeq 1.1\times 10^{-4}$~cm$^{3}$/mol-Mn$^{4+}$, $C\simeq 1.89$~cm$^{3}$K/mol-Mn$^{4+}$, and $\theta_{\rm CW}\simeq 56.6$~K. From the value of $C$ the effective magnetic moment is calculated using the relation $\mu_{\rm eff}= \sqrt{(3k_{\rm B}C/N_{\rm A})}\mu_{\rm B}$ to be $\mu_{\rm eff} \simeq 3.88$~$\mu_{\rm B}$, where $N_{\rm A}$ is the Avogadro's number, $k_{\rm B}$ is the Boltzmann constant, and $\mu_{\rm B}$ is the Bohr magneton. For a spin-3/2 system, the spin-only effective moment is expected to be $\mu_{\rm eff} = g\sqrt{S(S+1)} = 3.87~\mu_{\rm B}$, assuming Land$\acute{e}$ $g$-factor $g = 2$. Thus, our experimentally calculated $\mu_{\rm eff}$ value is very close to the expected value. 

Positive $\theta_{\rm CW}$ indicates dominant ferromagnetic interactions between the Mn$^{4+}$ ions. The core diamagnetic susceptibility $\chi_{\rm core}$ of LMMO was calculated to be $-1.24\times 10^{-4}$~cm$^3$/mol by adding the contributions of individual ions Li$^+$, Mg$^{2+}$, Mn$^{4+}$, and O$^{2-}$~\cite{Yogi024413,Hartree1130}. The Van-Vleck paramagnetic susceptibility, which mainly arises from second-order correction of Zeeman interaction in the presence of a magnetic field, was obtained by subtracting $\chi_{\rm core}$ from $\chi_0$ to be $\chi_{\rm VV} \simeq 4.52\times 10^{-4}$~cm$^3$/mol.

Inset of Fig.~\ref{Fig4}(b) presents a complete magnetic isotherm measured at $T = 1.8$~K. $M$ increases rapidly with $H$ in low fields and then saturates with $M_{\rm S} \simeq 3$~$\mu_{\rm B}/$Mn$^{4+}$, which is consistent with the expected value for spin-3/2 with $g=2$ ($M_{\rm S} = g S \mu_{\rm B}$). Further, no hysteresis is observed in the low-field region. All these features indicate that LMMO is a soft ferromagnet~\cite{Singh022506}.

\subsection{Heat Capacity}
\begin{figure}
\includegraphics[width=\columnwidth]{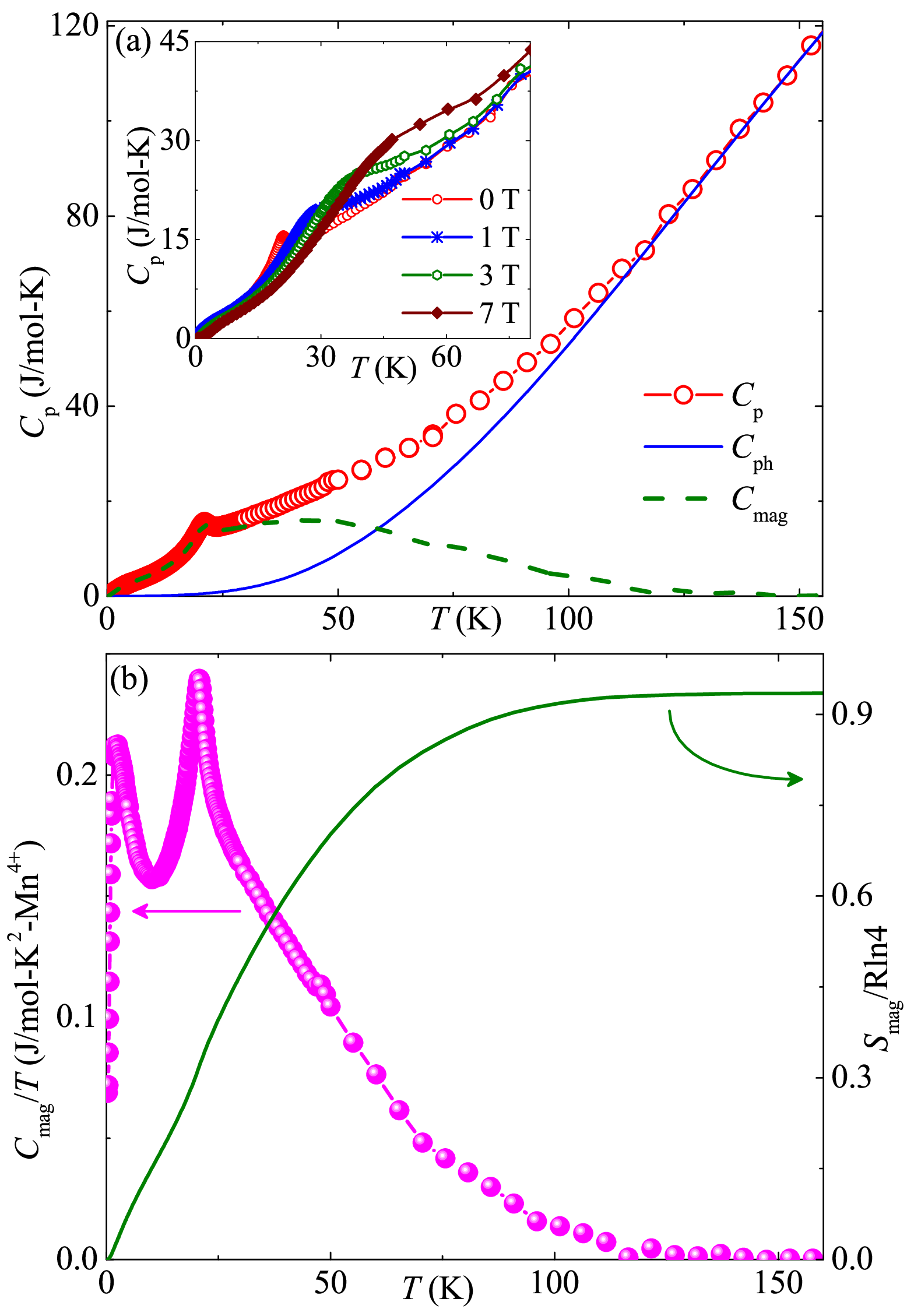}
\caption{\label{Fig5} (a) Heat capacity ($C_{\rm p}$) vs $T$ measured in zero-field. The solid line represents the phonon contribution to the heat capacity ($C_{\rm ph}$) determined using the Debye-Einstein fit and the dashed line represents the magnetic heat capacity ($C_{\rm mag}$). Inset: $C_{\rm p}(T)$ in different applied magnetic fields in the low-temperature regime. (b) $C_{\rm mag}/T$ and normalized magnetic entropy ($S_{\rm mag}/R\ln4$) as a function of temperature in the left and right $y$-axes, respectively.}
\end{figure}
Figure~\ref{Fig5}(a) displays the temperature-dependent heat capacity [$C_{\rm p}(T)$] down to 0.4~K in zero applied field. As the temperature decreases, $C_{\rm p}$ decreases systematically, showing a $\lambda$-type anomaly at around $T_{\rm C} \simeq 20.6$~K, indicating the onset of a magnetic long-range-order (LRO). Typically for a magnetic insulator, total heat capacity comprises two main contributions: phonon/lattice contribution ($C_{\rm ph}$) that is dominant in the high-temperature regime, and magnetic contribution ($C_{\rm mag}$) that exceeds $C_{\rm ph}$ at low temperatures.

In order to extract $C_{\rm mag}$ from the total heat capacity $C_{\rm p}$, we estimated the phonon contribution by fitting the high-temperature data using a linear combination of one Debye and three Einstein terms as~\cite{Gopal2012,Sebastian064413}
\begin{equation}\label{eq4}
C_{\rm ph} = f_{\rm D}C_{\rm D}(\theta_{\rm D}, T)+\sum_{i=1}^{3}g_{i}C_{Ei}(\theta_{Ei}, T).
\end{equation}
The first term in the above equation is the Debye model
\begin{equation}
C_{\rm D}(\theta_{\rm D},T)=9nR\left(\frac{T}{\theta_{\rm D}} \right)^3 \int_{0}^{\theta_{\rm D}/T}\frac {x^4e^x}{(e^x-1)^2} \,dx, \
\end{equation}
where $x=\frac{\hbar\omega}{k_{\rm B}T}$, $\omega$ is the vibration frequency, $R$ is the universal gas constant, and $\theta_{\rm D}$ is the characteristic Debye temperature. The high-energy vibration modes (optical phonons) are taken into account by the Einstein term,
\begin{equation}
C_{E}(\theta_{\rm E},T) = 3nR \left(\frac{\theta_{\rm E}}{T} \right)^2 \frac{e^{{\theta_{\rm E}}/T}}{(e^{\theta_{\rm E}/T}-1)^2},
\end{equation}
where $\theta_{\rm E}$ is the characteristic Einstein temperature. The coefficients $f_{\rm D}$, $g_{1}$, $g_{2}$, and $g_{3}$ are the weight factors that correspond to the number of atoms per formula unit ($n$). The fit of the zero-field $C_{\rm p}(T)$ data in the high-temperature region [see blue line in Fig.~\ref{Fig5}(a)] returns the characteristic temperatures: $\theta_{\rm D} \simeq 320$~K, $\theta_{\rm E1}\simeq 330$~K, $\theta_{\rm E2}\simeq 620$~K, and $\theta_{\rm E3}\simeq 1100$~K with $f_{\rm D}\simeq 0.071$, $g_{1}\simeq 0.143$, $g_{2}\simeq 0.571$, and $g_{3}\simeq 0.210$. One may notice that the sum of $f_{\rm D}$, $g_{1}$, $g_{2}$, and $g_{3}$ is close to one, as expected. Finally, the high-temperature fit was extrapolated down to low temperatures and subtracted from the experimental $C_{\rm p}(T)$ data to get $C_{\rm mag}(T)$. $C_{\rm mag}(T)/T$ vs $T$ is presented in the left $y$-axis of Fig.~\ref{Fig5}(b). The change in the magnetic entropy ($S_{\rm mag}$) is evaluated by integrating $C_{\rm mag}(T)/T$ [i.e. $S_{\rm mag} = \int_{0}^{T} \frac{C_{\rm mag}(T^{\prime})}{T^{\prime}}dT^{\prime}$] over the  entire temperature range. The value of $S_{\rm mag}$ is found to be $\sim 10.89$~J/mol-K which is close to $S_{\rm mag} = R\ln(2S+1)= 11.54$~J/mol-K, expected for a spin-$3/2$ system.

Another broad hump appears in the $C_{\rm p}$ data at around 2.5~K well below $T_{\rm C}$ which is more pronounced in the $C_{\rm mag}/T$ vs $T$ plot. However, it seems unlikely that another transition would occur within the ferromagnetic state. This second hump can be ascribed to the combined effect of change in the population of the Zeeman levels and energies of those levels arising from the $T$-dependent exchange field, typically expected for systems with large magnetic moments~\cite{Nath024431,Johnston094445}. The inset of Fig.~\ref{Fig5}(a) presents $C_{\rm p}(T)$ measured in different applied fields. The effect of magnetic field is clearly visible in the data. With the application of field, the peak at $T_{\rm C}$ broadens and shifts towards high temperatures, as expected in ferromagnets.


\subsection{Microscopic Modeling}
Table~\ref{tab:couplings} lists exchange couplings calculated for LMMO and for the isostructural compound Zn$_2$Mn$_3$O$_8$~\cite{kitani2021} along with the Curie-Weiss temperatures estimated using the mean-field expression
\begin{equation}
 \theta_{\rm CW}=\frac{S(S+1)}{3}\sum_i z_iJ_i,
\end{equation}
where $z_i$ is the number of couplings $J_i$ at a given Mn site ($z_1=z_2=4$, whereas $z_i=4$ for all other couplings). Our calculations perfectly reproduce the experimental Curie-Weiss temperatures of 56.6\,K (LMMO) and $-54$\,K (Zn$_2$Mn$_3$O$_8$~\cite{kitani2021}). Increasing the $U_d$ value of DFT+$U$ enhances ferromagnetic couplings and leads to a systematic rise in both Curie-Weiss temperatures (the one for Zn$_2$Mn$_3$O$_8$ thus becomes lower in magnitude and eventually ferromagnetic) but produces the same drastic difference between the $\theta_{\rm CW}$ values of the two compounds.

\begin{table}
\caption{\label{tab:couplings}
Exchange couplings in LMMO and Zn$_2$Mn$_3$O$_8$ calculated by the DFT+$U$ mapping analysis using $U_d=2$\,eV and $J_d=1$\,eV.
}
\begin{ruledtabular}
\begin{tabular}{ccr@{\hspace{1cm}}cr}
  & \multicolumn{2}{l}{LMMO} & \multicolumn{2}{l}{Zn$_2$Mn$_3$O$_8$} \\
  & $d_i$ (\r A) & $J_i$ (K) & $d_i$ (\r A) & $J_i$ (K) \smallskip\\
  \hline
$J_1$  & 2.868   & 	$-13.0$  &   2.921  &  $-1.6$  \\
$J_2$  & 4.983   &  $-2.2$   &   5.047  &  $-0.6$  \\
$J_3$  & 4.984   &  $-1.7$   &   4.953  &   4.8    \\
$J_4$  & 5.076   &   1.7     &   4.977  &   6.6    \\
$J_5$  & 5.106   &   0.9     &   5.047  &   2.7    \\
$J_6$  & 5.737   &   5.9     &   5.841  &   5.5    \\
$J_7$  & 5.790   &   0.0     &   5.801  &  $-1.4$  \\
$J_8$  & 5.843   &   3.5     &   5.760  &   6.6\smallskip\\
$\theta_{\rm CW}$ &  & 52  & &   $-51$
\end{tabular}
\end{ruledtabular}
\end{table}

An inspection of individual exchange couplings reveals the origin of this difference. LMMO is dominated by the ferromagnetic nearest-neighbor coupling $J_1$ on the hyperkagome lattice. This coupling is largely suppressed in Zn$_2$Mn$_3$O$_8$ and superseded by antiferromagnetic couplings beyond nearest neighbors. Such longer-range couplings are visibly enhanced in the Zn compound, thus explaining its antiferromagnetic behavior~\cite{kitani2021} vs the FM behavior of LMMO revealed in our work.

Our \textit{ab initio} results further put forward two effects that may be responsible for the reduced $T_{\rm C}$ of LMMO. First, FM nearest-neighbor couplings are frustrated by longer-range antiferromagnetic couplings that oppose ferromagnetic ordering. Second, the strength of exchange couplings in the Mn$^{4+}$ spinel compounds appears to be highly sensitive to the nonmagnetic cations, similar to the spin-$\frac12$ magnets of the $A$CuTe$_2$O$_6$ ($A$ = Sr, Ba, Pb) family~\cite{ahmed2015,bag2021,Chillal2348,chillal2021} that incidentally have the same structural symmetry. Residual Li/Mg disorder in LMMO~\cite{Branford1649} would modify exchange couplings in the vicinity of the antisite defects, resulting in an exchange randomness that should be further enhanced by a few percent of Mn occupying the Mg position according to Ref.~\cite{Branford1649}. This randomness may be another reason for the reduction in $T_{\rm C}$ and glassy dynamics observed in our work.

\subsection{Critical Analysis of Magnetization}
\begin{figure*}
\includegraphics[width=\textwidth]{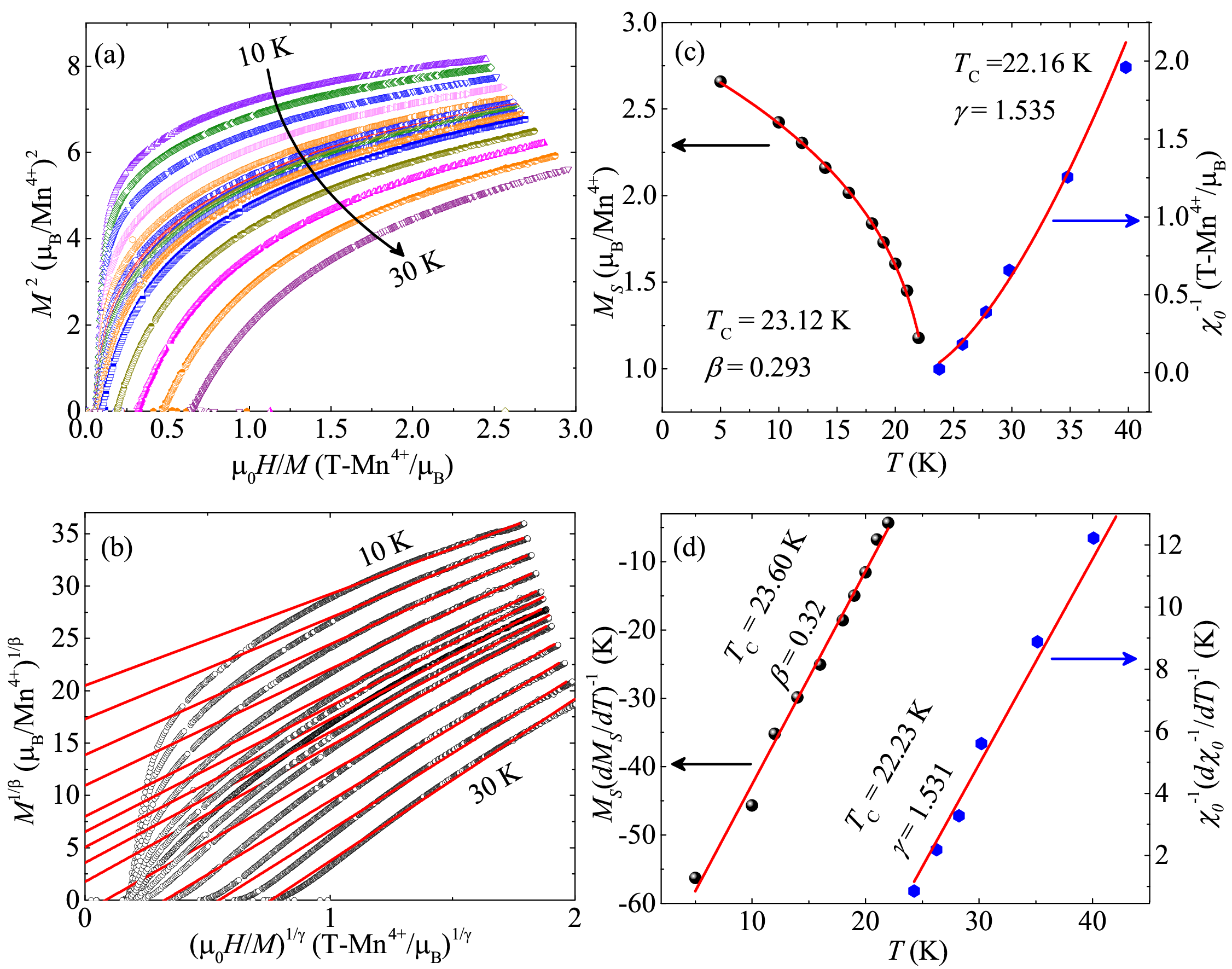}
\caption{\label{Fig6} (a) The Arrott plots ($M^{2}$ vs $H/M$) for LMMO at different temperatures around $T_{\rm C}$, (b) The modified Arrott plots [$M^{1/\beta}$ vs $(H/M)^{1/\gamma}$] at different temperatures, below and above $T_{\rm C}$. The solid lines are the linear fits to the data in the high-field regime. (c) Spontaneous magnetization $M_{\rm S}$ and zero-field inverse susceptibility $\chi_{0}^{-1}$ as a function of temperature in the left and right $y$-axes, respectively, obtained from the intercepts of the MAP in the vicinity of $T_{\rm C}$. The solid lines are the fits, as described in the text. (d) The Kouvel-Fisher plots for $M_{\rm S}$ and $\chi_{0}^{-1}$. The solid lines are the fits using Eqs.~\eqref{KF_Ms} and~\eqref{KF_chi0}, respectively.}
\end{figure*}
\begin{figure}
	\includegraphics[width=\columnwidth]{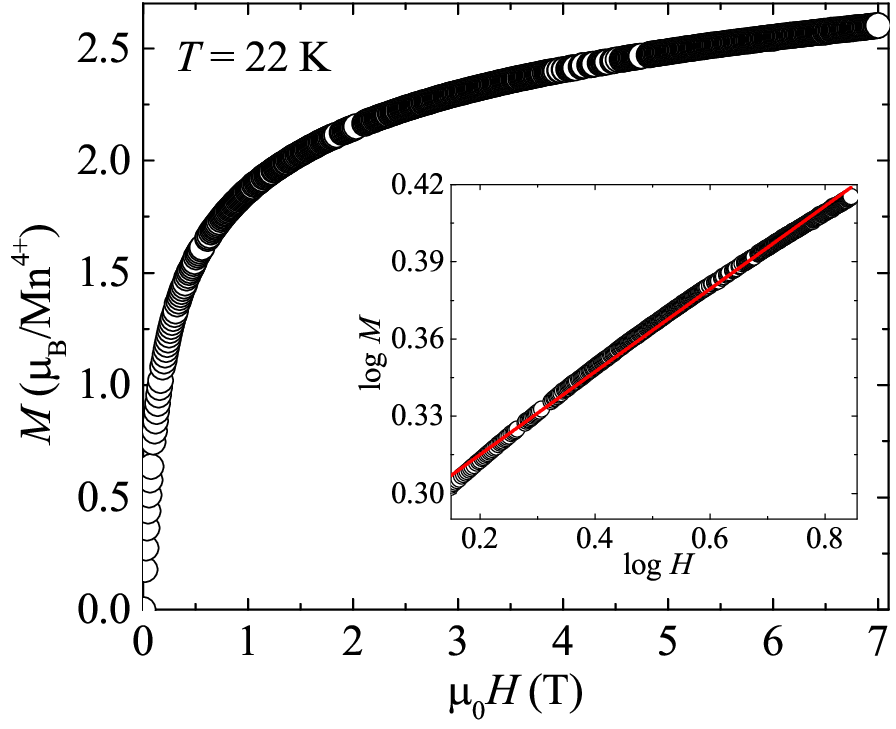}
	\caption{\label{Fig7} Magnetization isotherm ($M$ vs $H$) measured at $T = 22$~K ($\sim T_{\rm C}$). Inset: $\log M$ vs $\log H$ plot at $T = 22$~K.}
\end{figure}
The nature of the transition at $T_{\rm C}$ was further studied using Arrott plots of the magnetization~\cite{Arrott1394}. The Arrott plots assume the critical exponents to follow the mean-field model ($\beta = 0.5$, $\gamma = 1$). According to this method, the magnetization isotherms plotted in the form of $M^{2}$ vs $H/M$ produce a set of parallel lines around $T_{\rm C}$. Figure~\ref{Fig6}(a) shows the Arrott plot for LMMO near $T_{\rm C}$. All the curves in this plot reveal a non-linearity with the downward curvature even in the high-field regime, suggesting a non-mean-field type of behavior. According to the Banerjee criteria, a positive slope in $M^{2}$ vs $H/M$ curve indicates the second-order phase transition (SOPT), whereas a negative slope signifies the first-order transition~\cite{Banerjee16}. Thus, the observed positive slope in Arrott plots confirms the continuous SOPT in LMMO.

According to the scaling hypothesis, the universality class of the SOPT near $T_{\rm C}$ can be characterized by a set of the critical exponents ($\beta$, $\gamma$, and $\delta$) and magnetic equation of state~\cite{stanley1971}. Spontaneous magnetization ($M_{\rm s}$) at $T<T_{\rm c}$, zero-field inverse susceptibility ($\chi_{0}^{-1}$) at $T>T_{\rm c}$, and isothermal magnetization ($M$ vs $H$) at $T = T_{\rm c}$ are connected with the critical exponents by the following equations~\cite{Islam134433,Singh6981}:
\begin{equation}
\label{CA_Ms}
    M_{S} (T) = M_{0}(-\epsilon)^{\beta}~\text{for}~\epsilon < 0
\end{equation}
\begin{equation}
\label{CA_chi0}
    \chi_{0}^{-1} (T) = \Gamma(\epsilon)^{\gamma}~\text{for}~\epsilon > 0
\end{equation}
\begin{equation}\label{Cr_iso}
    M = XH^{1/\delta}~\text{for}~\epsilon = 0,
\end{equation}
where $\epsilon = \frac{T-T_{\rm c}}{T_{\rm c}}$ is the reduced temperature, while $M_{0}$, $\Gamma$, and $X$ are the critical amplitudes. The Arrott-Noakes equation of state can be written as~\cite{Arrott786},
\begin{equation}
(H/M)^{1/\gamma} = A\epsilon + BM^{1/\beta}.
\end{equation}
With the appropriate choice of $\beta$ and $\gamma$, the $M^{1/\beta}$ vs $(H/M)^{1/\gamma}$ plots [also known as the modified Arrott plot (MAP)] should produce a set of parallel lines in the high-field region for different temperatures around $T_{\rm C}$, whereas the isotherm at $T = T_{\rm C}$ should pass through the origin.

In this method, initial trial values of $\beta$ and $\gamma$ are taken from the 3D Heisenberg universality class, which produced more linear behavior in the high-field regime than the mean-field model. The linear fit to the MAP in the high-field region was extrapolated down to zero field and the values of $M_{S}(T)$ and $\chi_{0}^{-1}(T)$ were obtained from the intercepts of $M^{1/\beta}$ and $(H/M)^{1/\gamma}$ axes, receptively. The obtained temperature-dependent $M_{S}(T)$ and $\chi_{0}^{-1}(T)$ were fitted using Eqs.~\eqref{CA_Ms} and~\eqref{CA_chi0}, respectively, and the values of $\beta$ and $\gamma$ were estimated. This set of $\beta$ and $\gamma$ was again used to construct a new set of MAPs. This whole process was repeated several times until we got a set of parallel straight lines in the high-field regime with the stable values of $\beta$, $\gamma$, and $T_{\rm C}$. The final MAPs are shown in Fig.~\ref{Fig6}(b) and the obtained $M_{\rm S}$ and $\chi_{0}^{-1}$ as a function of temperature are depicted in Fig.~\ref{Fig6}(c). The fits using Eqs.~\eqref{CA_Ms} and \eqref{CA_chi0} yield ($\beta = 0.293$ with $T_{\rm C}=23.12$~K) and ($\gamma = 1.535$ with $T_{\rm C} = 22.16$~K), respectively. From Fig.~\ref{Fig6}(b) we noticed that the curves deviate from linearity in the low-field regime, which is due to the averaging over randomly oriented magnetic domains, typically observed in FM systems~\cite{Pramanik214426}.

To determine the critical exponents as well as $T_{\rm C}$ more precisely, we re-analyzed the data using the Kouvel-Fisher (KF) method~\cite{KouvelA1626}. The KF equations are
 \begin{equation}
 \label{KF_Ms}
 M_{S}(T) \left[\frac{dM_{S}(T)}{dT} \right]^{-1} = (1/\beta)(T-T_{\rm C})
\end{equation}
 and 
\begin{equation}
\label{KF_chi0}
\chi_{0}^{-1}(T) \left[\frac{d\chi_{0}^{-1}(T)}{dT} \right]^{-1} = (1/\gamma)(T-T_{\rm C}).
\end{equation}
 Here, the slope of the linear fits to $M_{S}(T)[\frac{dM_{S}(T)}{dT}]^{-1}$ vs $T$ and $\chi_{0}^{-1}(T)[\frac{d\chi_{0}^{-1}(T)}{dT}]^{-1}$ vs $T$ should yield $1/\beta$ and $1/\gamma$, respectively, whereas the intercept returns the $T_{\rm C}$. The KF plots for LMMO are shown in Fig.~\ref{Fig6}(d). The obtained critical exponents are $\beta = 0.32$ with $T_{\rm C} = 23.60$~K and $\gamma = 1.531$ with $T_{\rm C} = 22.23$~K. These values of the critical exponents match closely with those obtained from MAPs, suggesting a consistency between the two methods.

Following Eq.~\eqref{Cr_iso}, the $\log M$ vs $\log H$ plot at the critical temperature should produce a straight line with the slope of $1/\delta$. Figure~\ref{Fig7} presents the isotherm at $T\simeq T_{\rm C} = 22$~K. A straight-line fit (see inset of Fig.~\ref{Fig7}) in the log-log plot returns $\delta \simeq 6.18$. One can also estimate the value of $\delta$ from the Widom scaling relation, which connects the critical exponents $\beta, \gamma$, and $\delta$ in the following way~\cite{Widom3898}
\begin{equation}
\delta = 1 + \frac{\gamma}{\beta}.
\end{equation}
Using the values of $\beta$ and $\gamma$ from the MAPs and KF methods, $\delta$ is calculated to be $\sim 6.23$ and 5.78, respectively. These values are close to those obtained from the critical isotherm analysis. 

The critical exponents are compared with different universality classes in Table~\ref{CA_table}. The LMMO exponents do not fall under any universality class, but they are close to both 3D Heisenberg and 3D XY cases.

\begin{table*}
\caption{List of the critical exponents ($\beta$, $\gamma$, and $\delta$) and $T_{\rm C}$ of LMMO calculated from MAP, KF plot, critical isotherm
analysis, and MCE. Theoretical values for different universality classes are taken from Ref.~\cite{Islam134433}.}
\label{CA_table}
\begin{ruledtabular}
\begin{tabular}{c c c c c c c c c}
 Parameters & MAP &KF & Critical & MCE & Mean field & 3D Heisenberg & 3D XY & 3D Ising \\
  &&plot&isotherm & & model & model & model&model\\
 \hline
$\beta$&0.293&0.32&-- & &0.5&0.365&0.345&0.325\\
$\gamma$&1.535&1.531&-- & & 1&1.386&1.316&1.241\\
$\delta$&6.23&5.78&6.18 &6.25 & 3&4.8&4.8&4.82\\
$T_{\rm C}$~(K)&23&22.9&22&--& & & &\\
\end{tabular}
\end{ruledtabular}
\end{table*}

\subsection{Magnetocaloric Effect}
\begin{figure*}
\includegraphics[width=\textwidth]{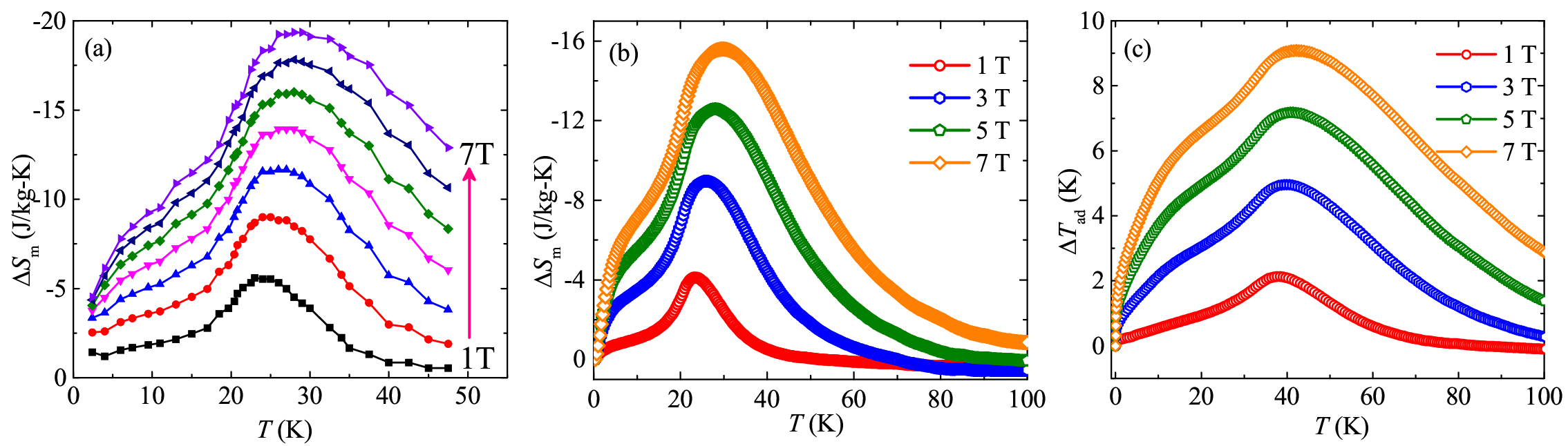}
\caption{\label{Fig8}(a) Isothermal magnetic entropy change ($\Delta S_{\rm m}$) as a function of $T$ calculated at various applied fields using Eq.~\eqref{eq13}. (b) Variation of $\Delta S_{\rm m}$ with $T$ calculated using field-dependent heat capacity data and Eq.~\eqref{eq14}. (c) Adiabatic temperature change ($\Delta T_{\rm ad}$) as a function of $T$ calculated at various applied fields using Eq.~\eqref{eq16}.}
\end{figure*}
Magnetocaloric effect (MCE) is defined as the change in temperature due to a change in the applied magnetic field. In order to achieve low temperatures, the magnetic field is first applied isothermally and then removed adiabatically. Therefore, MCE is generally quantified by the isothermal entropy change ($\Delta S_{\rm m}$) and adiabatic temperature change ($\Delta T_{\rm ad}$) with respect to the change in field ($\Delta H$). 

We calculated $\Delta S_{\rm m}$ from the magnetic isotherms ($M$ vs $H$) as well as from the field-dependent heat capacity data. To calculate MCE from the magnetic isotherms, the Maxwell thermodynamic relation $(\frac{\partial{S}}{\partial{H}})_{T}$ = $(\frac{\partial {M}}{\partial {T}})_{H}$ is utilized and $\Delta S_{\rm m}$ is estimated by integrating the equation as~\cite{Tishin2016},
\begin{equation}
\label{eq13}
\Delta S_{m}(H,T) = \int_{H_{i}}^{H_{f}}\frac {dM}{dT} \,dH.
\end{equation}
Figure~\ref{Fig8}(a) presents the variation of $\Delta S_{\rm m}$ as a function of temperature in different values of $\Delta H$ ($= H_{f} - H_{i}$). It manifests a maximum entropy change at around $T \simeq 27$~K, with the highest value of $\sim 20$~J/kg-K for a field change of $\Delta H = 7$~T.

Further, to cross-check the values of $\Delta S_{\rm m}$, we also estimated $\Delta S_{\rm m}$ using heat capacity data measured under zero and finite fields. First, we calculated the total entropy at field $H$ as 
\begin{equation}
\label{eq14}
S(T)_{H} = \int_{T_{i}}^{T_{f}}\frac {C_{\rm p}(T)_{H}}{T} \,dT, \
\end{equation}
where $C_{\rm p}(T)_{H}$ is the heat capacity measured under applied field $H$, and $T_{i}$ and $T_{f}$ represent the measured temperature range. In the next step, $\Delta S_{\rm m}$ was computed by taking the difference in entropy calculated in applied fields and in zero field i.e., $\Delta S_{\rm m}(T) = [S(T)_{H} - S(T)_{0}$].
Figure~\ref{Fig8}(b) depicts the variation of $\Delta S_{\rm m}(T)$ with temperature for different fields. The overall shape and peak position of the $\Delta S_{\rm m}$ curves are identical to the curves in Fig.~\ref{Fig8}(a), evaluated using the magnetic isotherms. A maximum value of $\Delta S_{\rm m}\simeq 16$~J/kg-K was obtained for the 7~T field. This further confirms the large value of $\Delta S_{\rm m}$ in LMMO.

We also calculated $\Delta T_{\rm ad}$ by two methods: first, from the combination of the zero-field heat capacity and magnetic isotherm data, and secondly by using only the heat capacity data measured in different applied fields. By the first method, $\Delta T_{\rm ad}$ is estimated using the relation~\cite{Akshata054076}
\begin{equation}
 \Delta T_{\rm ad} = -\int_{H_{i}}^{H_{f}}\frac{T}{C_{\rm p}}\frac {dM}{dT} \,dH.
\end{equation}
The maximum value of $\Delta T_{\rm ad}$ obtained by this method is around 35~K for $\Delta H = 7$~T (not shown here). The above equation can overestimate the value of $\Delta T_{\rm ad}$ because $T/C_{\rm p}$ is assumed to be constant over the whole range of applied fields~\cite{Pecharsky565}. Practically, there is a large change in $C_{\rm p}$ with magnetic field [see the inset of Fig.~\ref{Fig5}(a)] and this assumption fails in case of LMMO. Therefore, we try to estimate the value of $\Delta T_{\rm ad}$ by taking the difference between two temperatures corresponding to the same entropy but different fields as
\begin{equation}
\label{eq16}
\Delta T_{\rm ad} = T(S)_{H_{f}}-T(S)_{H_{i}}.
\end{equation}
The temperature variation of $\Delta T_{\rm ad}$ for different applied fields is shown in Fig.~\ref{Fig8}(c). The maximum value of  $\Delta T_{\rm ad}$ obtained using the above equation is around 9~K for 7~T which is significantly smaller than the value obtained using the former one. The latter method is more reliable as it considers the variation of $C_{\rm p}$ with field. Similar results are also reported for other FM compounds~\cite{Akshata054076,Pecharsky565}.

Relative cooling power ($RCP$) is another important parameter that determines the performance of a magnetocaloric material. It is defined as the amount of heat transfer between the hot and cold reservoirs in a refrigeration cycle. Mathematically, one can write it as
\begin{equation}
RCP = \int_{T_{\rm cold}}^{T_{\rm hot}}\Delta S_{m}(T,H) \,dT.
\end{equation}
Here, $T_{\rm cold}$ and $T_{\rm hot}$ are the temperatures of the hot and cold reservoirs, respectively. Thus, $RCP$ can be approximately determined as
\begin{equation}
|RCP|_{\rm approx} = \Delta S_{\rm m}^{\rm peak} \times \delta T_{\rm FWHM},
\end{equation}
where $\Delta S_{\rm m}^{\rm peak}$ is the peak value and $\delta T_{\rm FWHM}$ is the full width at half-maximum of the $\Delta S_{\rm m}$ vs $T$ plots in Fig.~\ref{Fig8}(a). The calculated value of $RCP$ for LMMO is about $\sim 840$~J/kg for an applied field of 7~T, which is significantly larger as compared to other magnetocaloric materials with $T_{\rm C}$ in the same temperature range (see Table~\ref{Table1}). Such a large value of $RCP$ in the case of LMMO is due to the distribution of entropy over a wide temperature range as clearly evident from Fig.~\ref{Fig8}(a). Indeed, a large MCE is anticipated in frustrated magnets~\cite{Tokiwa42}.

Generally, hydrogen gas liquefaction is necessary for efficient transport. In a recent report, the Carnot magnetic refrigerator (CMR) is proposed for highly efficient liquefaction by using magnetic cooling~\cite{Matsumoto012028} and the MCE materials with maximum entropy change around $20$~K are desired~\cite{Zhang092401}. Since the transition temperature ($T_{\rm C}\simeq 20.6$~K) falls within this range, LMMO seems to be an appropriate material for hydrogen liquefaction.

\begin{table}[h]
\caption{MCE parameters of LMMO are compared with previously studied materials having $T_{\rm C}$ (or $T_{\rm N}$) in the same temperature range. $\Delta H$ stands for the field change during the cooling cycle.}
\begin{tabular}{c c c c c c}
\hline\hline
 Compounds & $T_{\rm C}$ (or $T_{\rm N}$) & $|\Delta S_{\rm m}^{\rm peak}|$& $RCP$ & $\Delta{H}$ & Ref.  \\
     & (K)  & (J/kg-K) & (J/kg) & (T) \\
\hline
  PrCoB$_{2}$ & 18 & 8.1 & 104 & 5 & \cite{Li023903}   \\       
  GdCoB$_{2}$ & 25 & 17.1 & 462 & 5 & \cite{Li102509}  \\
  Gd$_{2}$NiSi$_{3}$ & 16.4 & 18.4 & 525 & 7 & \cite{Pakhira104414} \\
  ErFeSi & 22 & 23.1 & 365 & 5 & \cite{Zhang092401} \\
  LMMO & 20.6 & 20 & 840 & 7 & this work\\
  \hline\hline
\label{Table1}
\end{tabular}
\end{table}

\begin{figure}
\includegraphics[width=\columnwidth]{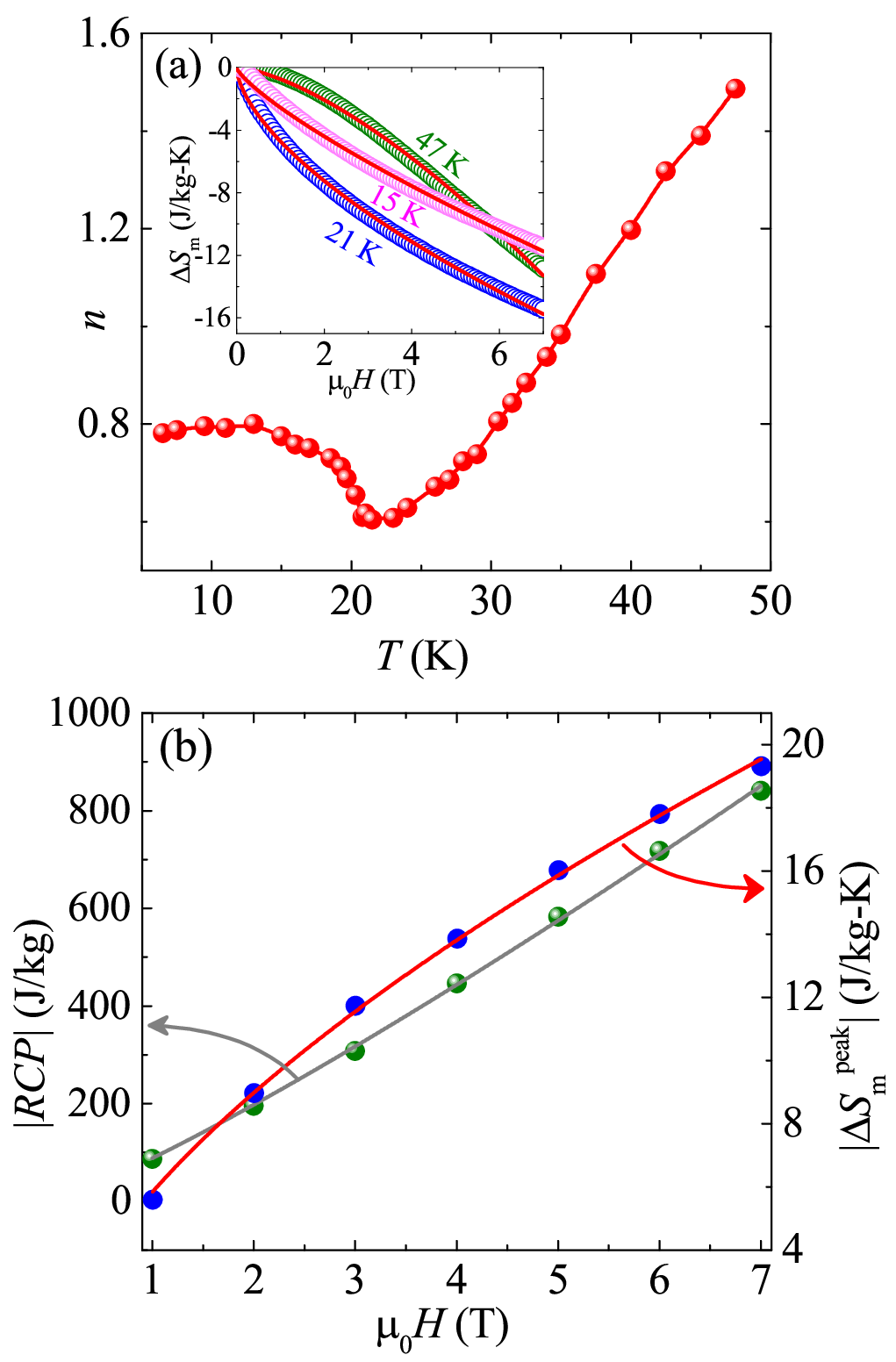}
\caption{\label{Fig9} (a) The exponent $n$ as a function of temperature obtained from fitting of $\Delta S_{\rm m}$ vs $H$ isotherms. Inset: $\Delta S_{\rm m}$ isotherms along with the fits for the temperatures across the $T_{\rm C}$. (b) The relative cooling power ($RCP$) and the absolute value of magnetic entropy change at the peak position ($\Delta S_{\rm m}^{\rm peak}$) as a function of magnetic field are plotted in the left and right $y$-axes, respectively. Solid lines are the power-law fits as described in the text.}
\end{figure}
The nature of the magnetic phase transition can be assessed from the MCE data. Large values of $\Delta S_{m}$ obtained for materials with first-order phase transitions are usually accompanied by relatively small $RCP$ values. Further, such materials also suffer from energy loss due to hysteresis that limits their usage in cyclic operations~\cite{Franco305}.
Therefore, the materials with SOPT are preferable for magnetic cooling.
To scrutinize the nature of the phase transition in LMMO, we fitted the $\Delta S_{\rm m}$ vs $H$ data at different temperatures near $T_{\rm C}$ by the power law $\Delta S_{\rm m}\propto H^{n}$ [see the inset of Fig.~\ref{Fig9}(a)]. The resulting exponent $n$ is plotted as a function of $T$ in Fig.~\ref{Fig9}(a). The $n$ values larger than 2 and smaller than 2 are expected for the first-order and second-order transitions, respectively~\cite{Law2680}. As shown in Fig.~\ref{Fig9}(a), $n(T)$ remains below 2 in the entire measured temperature range, thus confirming the SOPT in LMMO. In summary, LMMO satisfies most of the criteria for a practical MCE material: second-order nature of the transition, absence of hysteresis, and broad and asymmetric $\Delta S_{\rm m}$ curves resulting in large $RCP$ values. In Table~\ref{Table1}, we juxtapose LMMO with MCE candidates having $T_{\rm C}$ in the same range. Indeed, LMMO proves to be an excellent MCE material with the large $\Delta S_{\rm m}$ and $RCP$ values at low temperatures.

Moreover, one can also calculate the critical exponents by analyzing the MCE. To this end, we fitted the $RCP(H)$ and $\Delta S_{\rm m}^{\rm peak}(H)$ data by power laws of the form $RCP \propto H^{N}$ and $|\Delta S_{\rm m}^{\rm peak}| \propto H^n$, respectively. The fits shown in Fig.~\ref{Fig9}(b) return $n \simeq 0.62$ and $N \simeq 1.19$, respectively. These exponents ($n$ and $N$) are related to the critical exponents ($\beta$, $\gamma$, and $\delta$) as~\cite{Singh6981},
\begin{equation}\label{mce_n}
   n=1+\frac{\beta - 1}{\beta + \gamma} \text{ and }
   N=1+1/\delta.
\end{equation}
Using the values of the critical exponents ($\beta=0.293$, $\gamma=1.535$, and $\delta=6.23$) determined via critical analysis of the magnetization, we found $n \simeq 0.61$ and $N \simeq 1.16$ in a good agreement with the results of the MCE analysis.

\subsection{AC Susceptibility}
\begin{figure*}
	\includegraphics[width=\textwidth]{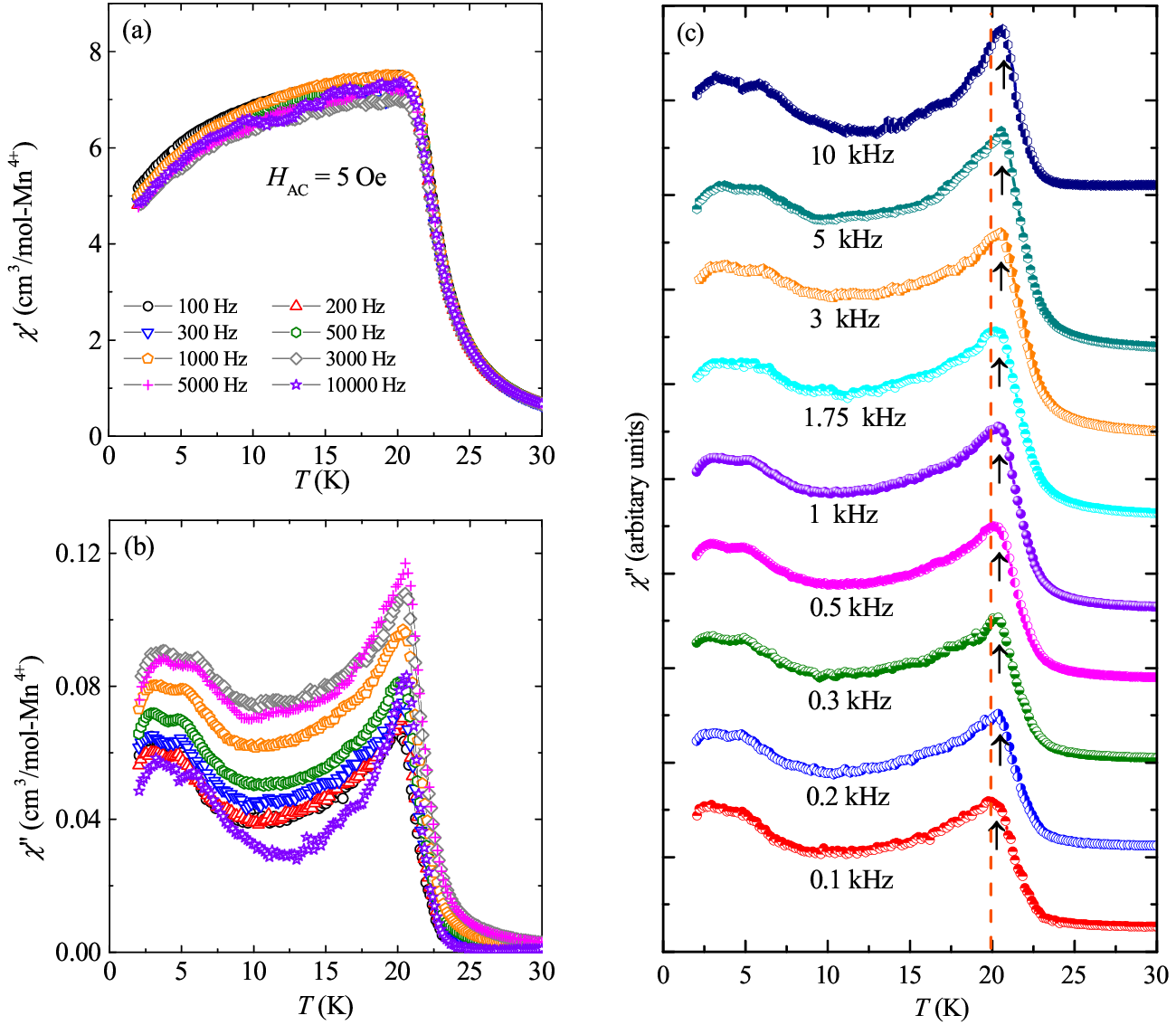}
	\caption{\label{Fig10} (a) Real part of the AC susceptibility ($\chi^{\prime}$) vs $T$ and (b) Imaginary part of the AC susceptibility ($\chi^{\prime\prime}$) vs $T$, measured at different frequencies. (c) Enlarged view of $\chi^{\prime\prime}$. The curves are vertically offset in order to highlight the shift of $T_{\rm f}$ with frequency.}
\end{figure*}
In order to shed further light on the bifurcation of the ZFC/FC susceptibilities near $T_{\rm C}$, we carried out AC susceptibility measurements at different frequencies with the fixed excitation field of $H_{\rm AC} = 5$~Oe. Figure~\ref{Fig10}(a) presents the temperature-dependent real part of the AC susceptibility ($\chi^{\prime}$) and shows an anomaly at $T_{\rm C}$, which shifts very weakly with frequency. However, the imaginary part of the AC susceptibility ($\chi^{\prime\prime}$) shows multiple anomalies at low temperatures, as depicted in Fig.~\ref{Fig10}(b). Among them, the anomaly near $T_{\rm C}$ is more distinct and frequency-dependent. This frequency dependence is additionally visualized in Fig.~\ref{Fig10}(c) that demonstrates the slight shift of the peak towards higher temperatures on increasing the frequency. This shift in the peak position signifies a glassy transition with the freezing temperature $T_{\rm f} \simeq 20$~K.

To characterize the nature of the glass transition in LMMO, we first calculate the Mydosh parameter ($\delta T_{\rm f}$) from the relative shift of $T_{\rm f}$ with respect to frequency,
\begin{equation}
\delta T_{\rm f} = \frac{\Delta T_{\rm f}}{T_{\rm f}\Delta \log_{10}\nu},
\end{equation}
where $\Delta T_{\rm f} = (T_{\rm f})_{\nu_1} - (T_{\rm f})_{\nu_2}$ and $\Delta\log_{10}(\nu)=\log_{10}(\nu_1)-\log_{10}(\nu_2)$. The difference is taken between the lowest and highest frequencies of $\nu_{1} = 100$~Hz and $\nu_{2} = 10$~kHz, respectively, resulting in $\delta T_{f} \simeq 0.015$ for LMMO. Generally, in canonical SG systems such as AuMn and CuMn the reported values are $\delta T_{f} \simeq 0.0045$~\cite{Mulder545} and $\sim 0.005$~\cite{Mydosh280}, whereas superparamagnetic system like LNMO feature much higher values on the order of $\delta T_{\rm f} \simeq 0.4$~\cite{Islam134433}. The intermediate value for LMMO is typical of cluster-glass systems like Cr$_{0.5}$Fe$_{0.5}$Ga ($\delta T_{\rm f}\simeq 0.017$)~\cite{Bag144436} and suggests the formation of the cluster-glass state in LMMO.

\begin{figure}
	\includegraphics[width=\columnwidth]{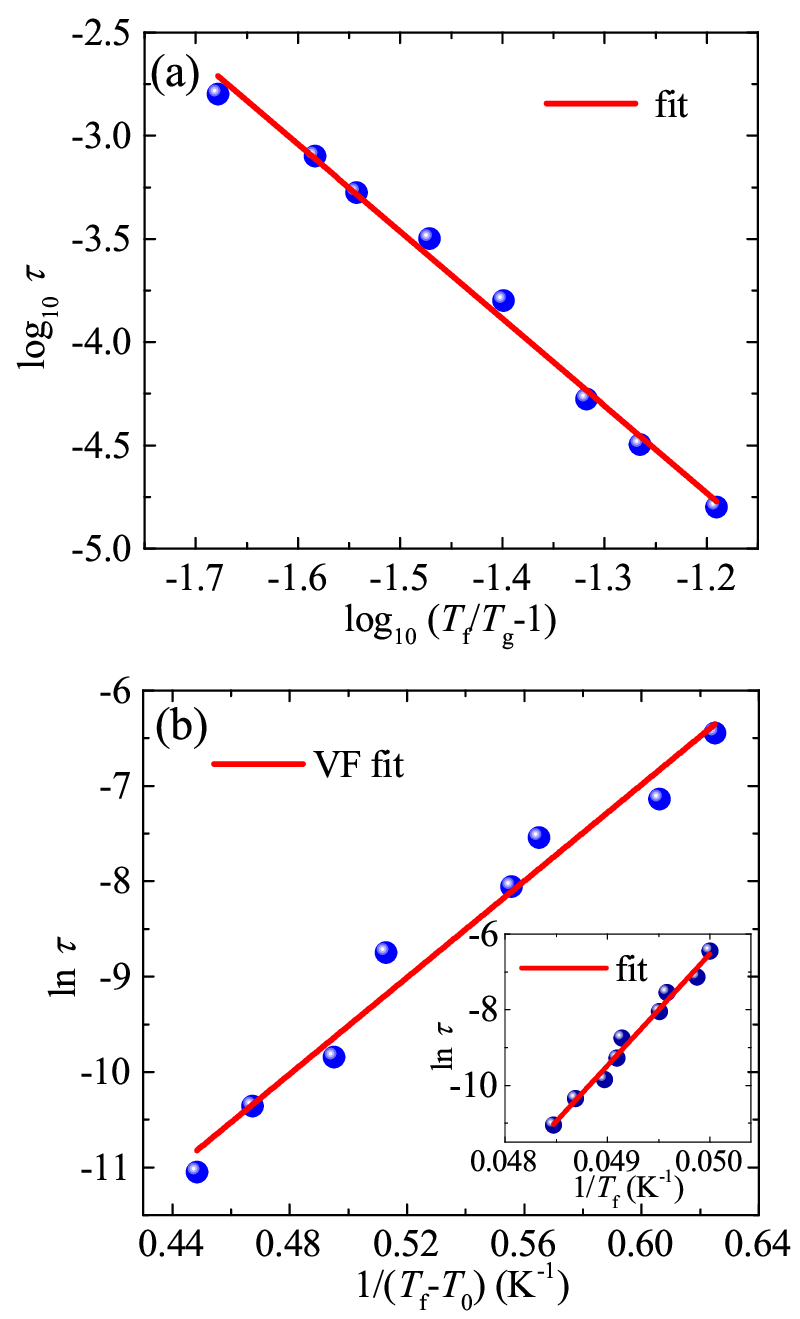}
	\caption{\label{Fig11} (a) $\log_{10}\tau$ vs $\log_{10}(T_{\rm f}/T_{\rm g}-1)$. The solid line is the fit using Eq.~\eqref{eq23}. (b) $\ln\tau$ vs $1/(T_{\rm f}-T_{0})$. The solid line represents the fit using Eq.~\eqref{eq26}. Inset: Arrhenius fit to $\ln \tau$ vs $1/T_{\rm f}$.}
\end{figure}
The shift of $T_{\rm f}$ with frequency can be described by a power law of the dynamical scaling theory as~\cite{Mydosh280,Hohenberg435}
\begin{equation}\label{eq22}
    \tau = \tau^{*}\left[\frac{T_{\rm f}-T_{\rm g}}{T_{\rm g}}\right]^{-z\nu^{\prime}}.
\end{equation}
Here, $\tau$ is the dynamical fluctuation time scale calculated as $\tau = 1/(2\pi\nu)$ and $\nu$ is the AC frequency. $\tau$ depends on $\tau^{*}$, which is the relaxation time for a single spin flip, $T_{\rm g}$ is the freezing temperature when $\nu$ approaches zero, $z$ is the dynamic critical exponent, and $\nu^{\prime}$ corresponds to the critical exponent of the correlation length $\zeta = (T_{\rm f}/T_{\rm g}-1)^{-\nu^{\prime}}$.
To determine $\tau^*$ and $z\nu^{\prime}$ from the fit, Eq.~\eqref{eq22} can be re-written as
\begin{equation}\label{eq23}
    \log_{10}\tau = \log_{10}\tau^{*}-z\nu^{\prime}\log_{10}(T_{\rm f}/T_{\rm g}-1).
\end{equation}
As shown in Fig.~\ref{Fig11}(a), we have plotted $\log_{10}\tau$ vs $\log_{10}(T_{\rm f}/T_{\rm g}-1)$ and tried to obtain the best fit using Eq.~\eqref{eq23}, fixing different values of $T_{\rm g}$. The final fit returns $z\nu^{\prime} \simeq 4.2\pm 0.2$ and $\tau^{*} \simeq (1.6 \pm 0.2)\times10^{-10}$~s for $T_{\rm g} \simeq 19.4$~K. For any SG system, the value of $z\nu^{\prime}$ varies between 4 and 12 and our experimental value clearly falls within this range~\cite{Bag144436,Rahul214427}. Further, a relatively large value of $\tau^{*}$ reflects that the spin dynamics is slower than in the conventional SG systems with $\tau^*\sim 10^{-13}$~s~\cite{Mydosh280}. One can differentiate between canonical SG and cluster-glass systems based on the value of $\tau^{*}$. For instance, for a canonical SG, $\tau^{*}$ can have a value between $\sim 10^{-12}$ and $10^{-13}$~s, whereas for cluster glass it is expected to lie in the range of $\sim 10^{-7}$ to $10^{-10}$~s~\cite{Pakhira104414,Anand014418}. Therefore, our value of $\tau^{*}$ gives further evidence for the cluster-glass state in LMMO~\cite{Bag144436}.

To shed light on the interactions between the magnetic entities, we fitted the frequency dependence of $T_{\rm f}$ using the Arrhenius law. This law is derived assuming negligible or weak interactions in the system. It has the form~\cite{Binder801}
\begin{equation}\label{eq24}
\tau= \tau_{0}\exp \left (\frac{E_{\rm a}}{k_{\rm B}T_{\rm f}} \right),
\end{equation}
where $\tau_{0}$ is the relaxation time for a single spin flip like $\tau^{*}$, and $\frac{E_{\rm a}}{k_{\rm B}}$ represents the average activation energy of the relaxation barrier. For fitting purpose, $\ln \tau$ vs $1/T_{\rm f}$ is plotted in the inset of Fig.~\ref{Fig11}(b). A linear fit returns $\tau_{0}\simeq 1\times10^{-67}$~s and $\frac{E_{\rm a}}{k_{\rm B}} = (2955\pm 62$)~K. Clearly, these values are unphysical implying the failure of the Arrhenius law. Therefore, the dynamics in LMMO can not be described with single spin flips, it must be cooperative in nature~\cite{Bag144436}.

The interactions between the dynamic entities can be taken into account using the Vogel-Fulcher (VF) law that introduces the ($T_{\rm f}-T_0$) term into the previous equation~\cite{Mydosh280,Souletie516},
\begin{equation}\label{eq25}
    \tau = \tau_{0}\exp \left[\frac{E_{\rm a}}{k_{\rm B}(T_{\rm f}-T_{0})}\right].
\end{equation}
Here, $T_{0}$ is the empirical VF temperature, which describes the interactions. To fit the data, we rewrote Eq.~\eqref{eq25} as
\begin{equation}\label{eq26}
    \ln\tau= \ln\tau_{0}+ \left[\frac{E_{\rm a}}{k_{\rm B}(T_{\rm f}-T_{0})} \right].
\end{equation}
The fit shown in Fig.~\ref{Fig11}(b) returns $T_0 \simeq 18.40$~K, $\tau_{0} = (2.4 \pm 0.9)\times 10^{-10}$~s, and $\frac{E_{\rm a}}{k_{\rm B}} \simeq (25 \pm 2)$~K. The non-zero value of $T_{0}$ confirms the formation of clusters, whereas $\frac{E_{\rm a}}{k_{\rm B}T_f}\sim 1.25$
suggests an intermediate coupling strength. The Tholence criterion $\delta T_{\rm h}=\frac{T_{\rm f}-T_{0}}{T_{0}}$ is also used to compare different glassy systems~\cite{Tholence157}. In our case, $\delta T_{\rm h}\sim 0.0869$ is comparable to other cluster-glass systems~\cite{Bag144436,Anand014418}. The agreement of the Tholence criteria further indicates the cluster-glass dynamics of LMMO.

\subsection{Magnetic Relaxation}
\begin{figure}
\includegraphics[width=\columnwidth]{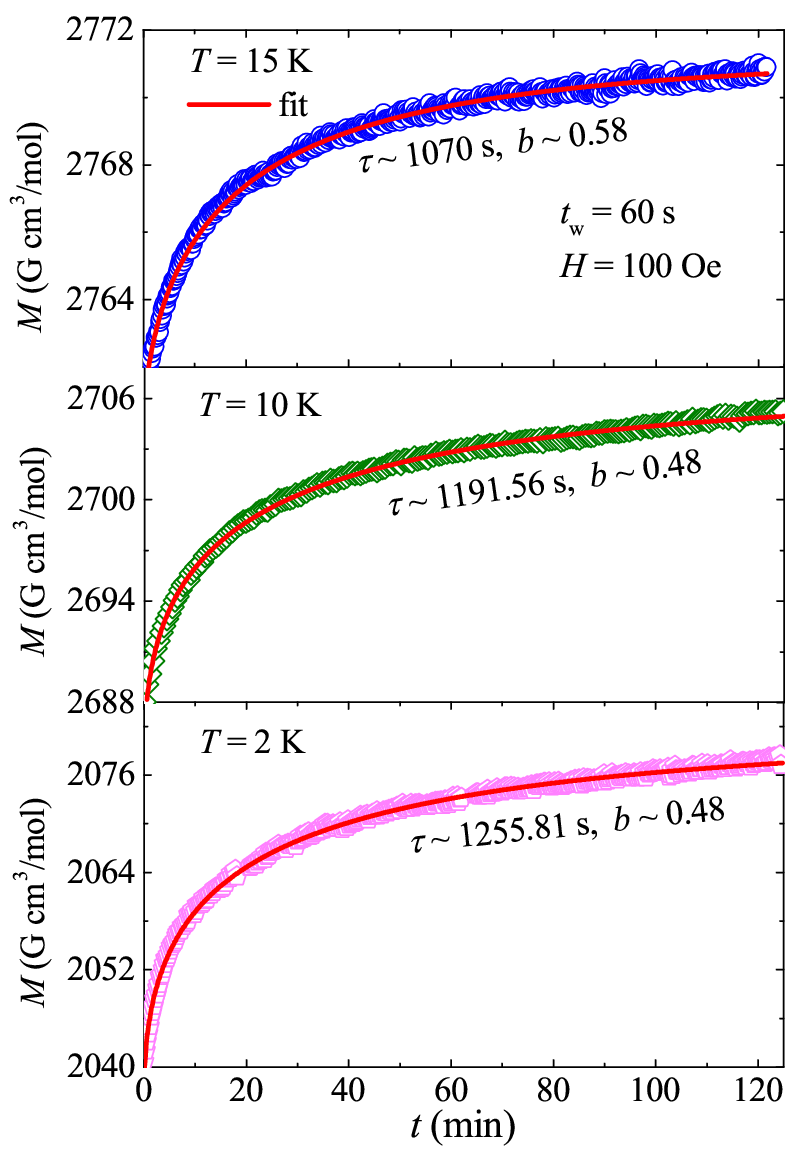}
\caption{\label{Fig12} Growth of magnetization as a function of time at different temperatures ($T= 2$~K, 10~K, and 15~K) in $H = 100$~Oe and after a waiting time of 60~s. The solid red lines represent the fits using Eq.~\eqref{eq27}.}
\end{figure}
Having confirmed the cluster-glass state in LMMO, we investigate the non-equilibrium dynamics and magnetic memory effect. Time dependence of the magnetization was measured 
at three different temperatures (15, 10, and 2~K) below $T_{\rm f}$. The sample was cooled down from the paramagnetic state ($T \ge 200$~K) to the required temperature in zero field. After a waiting time of $t_{\rm w} = 60$~s at that temperature, a small magnetic field of 100~Oe was applied and $M(t)$ was recorded as depicted in Fig.~\ref{Fig12}. This time dependence of the magnetization can be fitted using the standard stretch exponential function~\cite{Bag144436,Islam134433,Ulrich024416},
\begin{equation}\label{eq27}
    M(t) = M_{0} - M_{\rm g} e^{{- \left (\frac{t}{\tau} \right)^b}}.
\end{equation}
Here, $M_{0}$ is the magnetization at $t=0$, $M_{\rm g}$ represents the glassy component of the magnetization, $\tau$ is the characteristic time constant, and $b$ is the stretching exponent. $b$ is indicative of the spin dynamics because it strongly depends on the energy barrier involved in the relaxation process. For a glassy system, the typical value of $b$ lies between 0 and 1. In this relation, $b = 0$ implies $M(t)$ = constant i.e. no relaxation at all, whereas $b = 1$ implies that the system relaxes with a single time constant. Intermediate values of $b$ indicate nonuniform energy barriers of the spin relaxation. The $M(t)$ curves of LMMO are well-fitted using Eq.~\eqref{eq27}, resulting in $b \simeq 0.48$, $0.48$, and $0.58$ for $T = 2$, 10, and $15$~K, respectively. The obtained $b < 1$ implies the evolution of magnetization through a number of intermediate meta-stable states with non-uniform energy barriers. Indeed, a reduced value of $b$ ($b < 1$) is typically reported for many SG and superparamagnetic compounds~\cite{De033919, Johannes174410}. Further, the value of $\tau$ decreases with increasing temperature below $T_{\rm f}$ as expected for glassy systems~\cite{Bag144436,Ghara024413}.

\subsection{Magnetic Memory Effect}
\begin{figure}
	\includegraphics[width=\columnwidth]{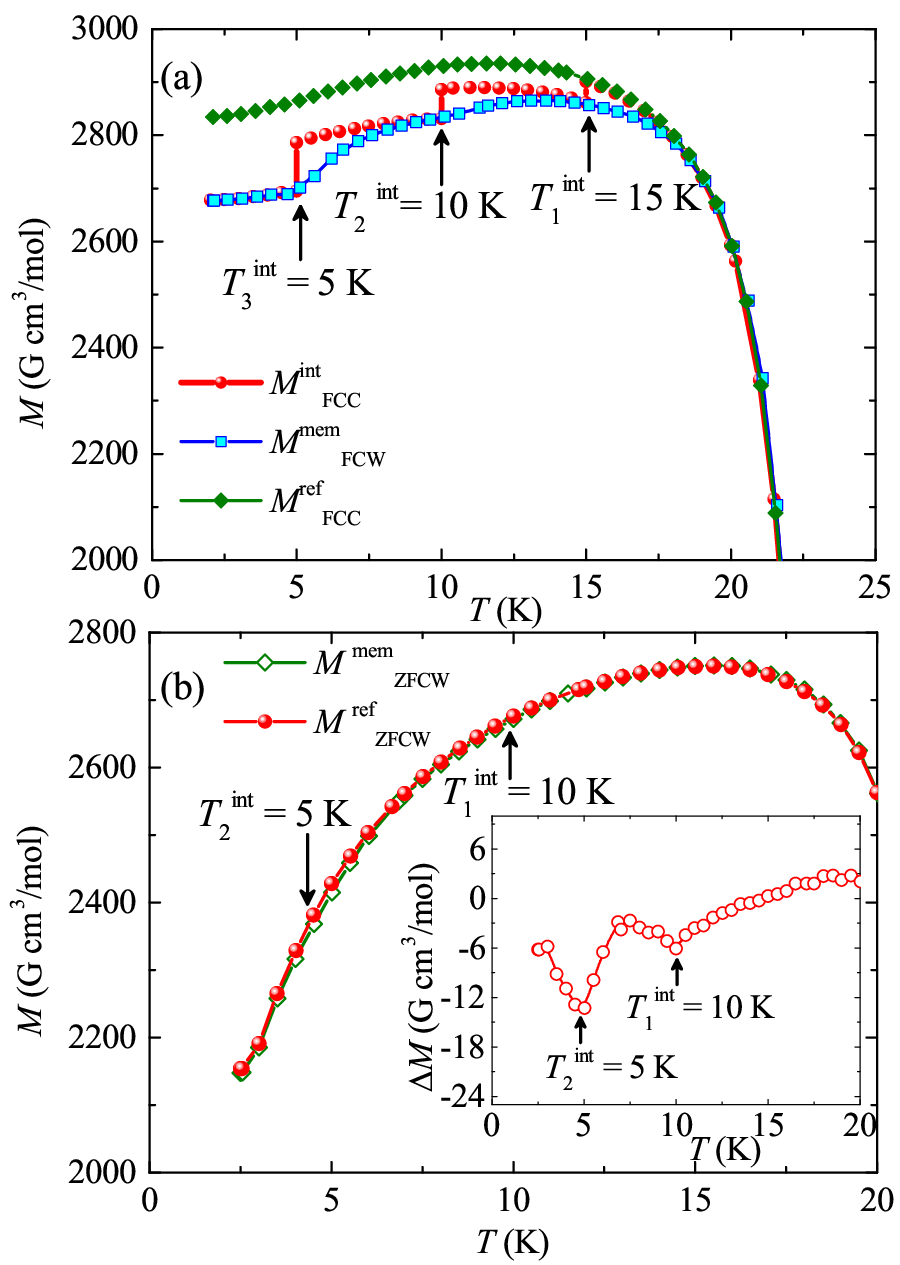}
	\caption{\label{Fig13} Memory effect as a function of temperature in (a) FC and (b) ZFC protocols in an applied field of $H = 100$~Oe, as discussed in the text. The measurements are interrupted at 15~K, 10~K, and 5~K for 2 hours each. Inset: Difference in magnetization $\Delta M = M_{\rm ZFCW}^{\rm mem} - M_{\rm ZFCW}^{\rm ref}$ vs $T$ for ZFC protocol.}
\end{figure}
The bifurcation of the $\chi(T)$ curves in ZFC and FC protocols is typically observed for SG as well as the SP systems. These scenarios can be distinguished by the presence of the magnetic memory effect. Superparamagnetic systems, which are non-interacting in nature, can show the FC memory only~\cite{De033919}, while interacting glassy systems may exhibit both FC and ZFC memory~\cite{Markovich134440}.

First, we will discuss the FC memory measurements (Fig.~\ref{Fig13}). In this protocol, we applied the 100 Oe field and cooled the sample from 200~K to the base temperature of 2~K with a constant slow cooling rate of 0.5~K/min. The cooling was not continuous but interrupted at three different temperatures $T_{1}^{\rm int} = 15$~K, $T_{2}^{\rm int} = 10$~K, and $T_{3}^{\rm int} = 5$~K, which are below $T_{\rm f}$ for a period of $t_{\rm w} = 2$ hrs each. In this waiting period, the magnetic field was switched off in order to allow the system to relax. After each $t_{\rm w}$, the magnetic field was again switched on and the field cooled cooling (FCC) process was resumed. The magnetization recorded during this process is denoted as $M_{\rm FCC}^{\rm int}$. As shown in Fig.~\ref{Fig13}(a), step-like features at 15~K, 10~K, and 5~K are observed. Once it reached 2~K, the sample was heated with the same slow rate of 0.5~K/min, and its magnetization was recorded in the same magnetic field of 100~Oe without any interruption till 200~K, referred as $M_{\rm FCW}^{\rm mem}$ (FCW stands for field cooled warming). Interestingly, $M_{\rm FCW}^{\rm mem}$ shows a slope change at all the three interruption temperatures. This clearly suggests that the system remembers its thermal history of magnetization during the cooling, thus showing the FC memory. A cooling curve is also measured in the same field without any interruption for reference ($M_{\rm FCC}^{\rm ref}$).


Similarly, we have measured the memory effect in the ZFC protocol. 
The $M_{\rm ZFCW}^{\rm mem}$ (ZFCW stands for zero field cooled warming) and $M_{\rm ZFCW}^{\rm ref}$ curves are found to overlap except at the interruption temperature regions. The difference in the magnetization ($\Delta M = M_{\rm ZFCW}^{\rm mem} - M_{\rm ZFCW}^{\rm ref}$) is shown in the inset of Fig.~\ref{Fig13}. It manifests the memory dips at 5~K and 10~K. The absence of the memory dip at 15~K may be because 15~K is just below $T_{\rm f}\sim 20$~K and the memory effect is not pronounced~\cite{Rahul214427}. However, the dips at two temperatures well below $T_{\rm f}$ confirm the ZFC memory of the system.

The observation of both FC and ZFC memories further confirms the cluster-glass state in LMMO. The ZFC memory of the interacting glassy systems is elucidated by different theoretical models, one being the random energy model~\cite{Derrida2613}. According to this model, below $T_{\rm f}$, a large number of equally probable states exists with a random local mean dipolar field. In the normal state, the local energy barriers between these states are low. On cooling the system in zero field and allowing it to relax at different temperatures below $T_{\rm f}$, the energy barriers become higher as time progresses. This increase in the barrier height is more pronounced at lower temperatures. Next, when we apply the field and do the measurements under ZFCW condition, the system fails to recover the total magnetization, and a dip is observed at each stopping temperature~\cite{Bandyopadhyay214410}. This also explains why the dip is more pronounced at low temperatures. On the other hand, in non-interacting systems like superparamagnets, no ZFC memory is observed because there exist only two states (up and down) with equal probability~\cite{Sasaki104405}.

\subsection{Memory Effect using Magnetic Relaxation}
\begin{figure}
\includegraphics[width=\columnwidth]{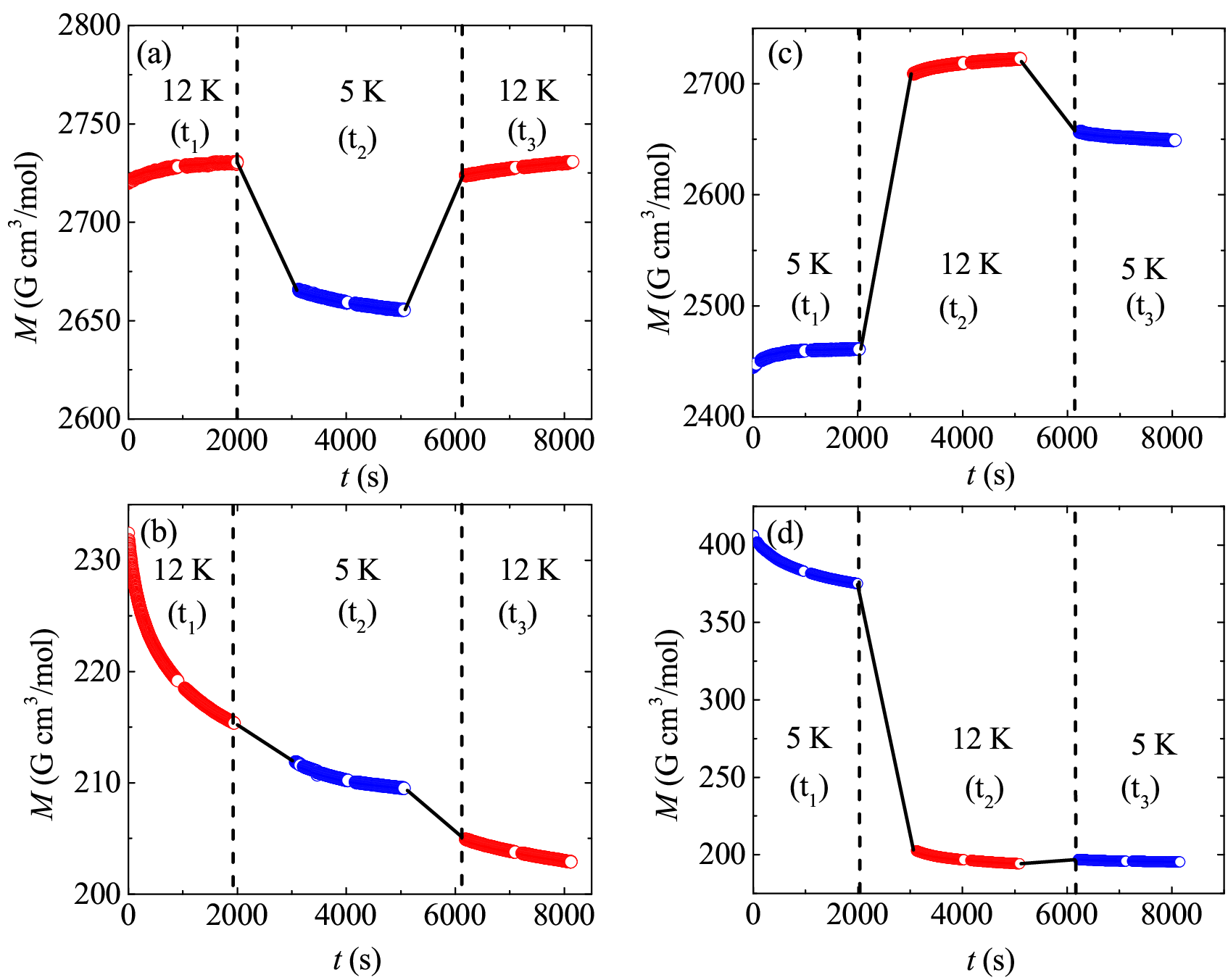}
\caption{\label{Fig14} Negative-$T$ cycle measured in an applied field of 100~Oe for both (a) ZFC and (b) FC processes. For the positive-$T$ cycle, ZFC and FC data are shown in (c) and (d), respectively.}
\end{figure}
To understand the mechanism behind the magnetic memory exhibited by LMMO and the influence of temperature cycling, we investigated the magnetic relaxation following the protocol of Sun~\textit{et. al.}~\cite{Sun167206} for both positive and negative-$T$ cycles.

$Negative-T~cycle$: Figures~\ref{Fig14}(a) and (b) present the behavior of magnetic relaxation recorded for the negative-$T$ cycle in ZFC and FC protocols, respectively. In the ZFC mode, the sample was cooled down from 200~K to 12~K (below $T_{\rm f}$) in zero field. At 12~K, a small field of 100~Oe was applied, and $M(t)$ was recorded for almost $t_{1}\simeq 1$~hr. It was found to increase exponentially with $t$. After that, the sample was cooled down to a lower temperature of 5~K in the same magnetic field, and again $M(t)$ was measured for a period of $t_{2}\simeq 1$~hr. The nature of the $M(t)$ curve was found to be nearly constant or very weakly $t$-dependent. Subsequently, the temperature was restored back to 12~K under the same magnetic field and $M(t)$ was measured again for $t_{3}\simeq 1$~hr. It was found to grow exponentially with $t$. In the FC process, the sample was cooled from 200~K to 12~K in a small applied field of 100~Oe. Once it reached 12~K, the magnetic field was switched off, and $M(t)$ was measured for $t_{1}\simeq 1$~hr, which is found to decay exponentially with $t$. Then, the sample was cooled down to 5~K in zero field, and $M(t)$ was recorded for $t_{2}\simeq 1$~hr, which shows a weak $t$-dependence. Finally, the system was warmed back to 12~K and $M(t)$ was again measured for $t_{3}\simeq 1$~hr in zero field, which is found to decay exponentially with $t$ as observed during $t_{1}$ but with a small offset. Generally, for a glassy system, one expects $M$ to be $t$-independent during $t_2$ and the magnetization during $t_{1}$ and $t_{3}$ when put together (skipping $M$ during $t_2$) should exhibit a continuous growth or decay with $t$ for both ZFC and FC protocols~\cite{Bag144436}. However, in the present case, the measured $M$ during $t_2$ is not strictly $t$-independent and there is a small offset between the data measured during $t_1$ and $t_3$. This offset is more distinct for the FC protocol. This suggests that during the -ve cycle some metal-stable states might have developed with low energy barriers. These states possibly relax during $t_{2}$ even at low temperatures (5~K), resulting in a weak $t$-dependence. Therefore, when the system is heated back to $12$~K, it fails to achieve the original magnetization value, hence an offset. This clearly suggests the occurrence of multiple relaxation processes in LMMO. Similar results are indeed reported in Ref.~\cite{Majumder024408}. Nevertheless, $M$ follows an exponential behaviour during both $t_1$ and $t_3$. Thus, the negative-$T$ cycle measured under ZFC and FC modes still preserves the memory (though not entirely) even after a temporary cooling. This is a simple demonstration of the magnetic memory effect.

$Positive-T~cycle$: Similar to the negative-$T$ cycle, we also recorded magnetization relaxation [$M(t)$] for the positive-$T$ cycle in both ZFC and FC protocols, which are depicted in Fig.~\ref{Fig14}(c) and (d), respectively. In the ZFC process, the sample was cooled down from 200~K to 5~K in zero field. At 5~K, a small magnetic field of 100~Oe was applied and magnetization was recorded for $t_{1}\simeq 1$~hr. It is found to grow with $t$. Then, the sample was heated up to 12~K in the same applied field and $M(t)$ was measured for $t_{2}\simeq 1$~hr, which also shows the increasing behavior with $t$. Finally, the system was brought back to 5~K, but the recorded $M(t)$ showed a constant behavior with $t$. In the FC process, the system was cooled down to 5~K from 200~K in a small applied field of 100~Oe. Once it reached 5~K, the magnetic field was switched off, and $M(t)$ was recorded with the same sequence as for the ZFC process. The obtained results are shown in Fig.~\ref{Fig14}(d). It follows the same trend but in the opposite way as for the ZFC protocol. Unlike the negative-$T$ cycle, there is no continuity in the recorded $M(t)$ during $t_{1}$ and $t_{3}$ at 5~K. This clearly suggests that the positive-$T$ cycle erases memory and rejuvenates the magnetic relaxation process. That is why no memory effect is observed when the temperature is restored back to 5~K.

Two theoretical models are commonly used to understand the relaxation process of disordered systems. One is the droplet model~\cite{Fisher373}, which supports the symmetric response in the magnetic relaxation process in both negative and positive cycles. Another one is the hierarchical model~\cite{Lefloch647} that predicts an asymmetric response in these cycles. The hierarchical model is applied to a system having a multi-valley free energy landscape, whereas for the droplet model only one spin configuration is favored. In our system, the presence and absence of memory effect in the negative and positive $T$-cycles, respectively, imply an asymmetric behavior and support the hierarchical model proposed for the disordered systems. Basically, in the hierarchical model, a multi-valley free-energy surface exists at a given temperature. When we cool the system from $T$ to $T-\Delta T$, each valley splits into many sub-valleys. If $\Delta T$ is large, the energy gap between the primary valleys becomes high and the system fails to overcome these energy barriers within a finite waiting time $t_2$. Therefore, the relaxation process occurs within the sub-valleys or metastable states. When the temperature of the system is reinstated to the original temperature, the sub-valleys merge back to their initial free-energy surface, and the relaxation at $T$ resumes without being perturbed by the intermediate relaxations at $T-\Delta T$. However, when the temperature of the system is increased from $T$ to $T+\Delta T$, then the energy barriers between the primary valleys are very low or sometimes they even get merged. Therefore, the relaxation occurs between different valleys. When the temperature of the system is brought back to the initial value $T$, the relative occupancy of each valley does not remain the same as before, even though the free-energy surface is back to its original state. Thus, the state of the system changes after a positive cycle and shows no memory effect~\cite{Lefloch647}. Since the hierarchical organization of the metastable states requires coupling among a large number of degrees of freedom, the observed behavior implies the interaction among the dynamic entities. Thus, this further supports the cluster-glass formation in LMMO~\cite{Sun167206}.

\section{Summary}
In summary, we have successfully synthesized and report the physical properties of a new frustrated hyperkagome compound LMMO. No signature of structural transition is found in temperature-dependent XRD measured down to 13~K. The thermodynamic measurements suggest the onset of FM correlations at $T_{\rm C} \simeq 20.6$~K. The CW temperature ($\theta_{\rm CW} \simeq 56.6$~K) also implies dominant FM interaction, setting a frustration parameter $f = |\theta_{\rm CW}|/T_{\rm C} \simeq 2.8$.

A detailed critical analysis of magnetization data is carried out using modified Arrott plot and Kouvel Fisher methods and the critical exponents are estimated. These exponent values are reproduced via various analysis methods and the Widom scaling relation, indicating the robustness of the critical analysis technique. Though the obtained values of critical exponents do not exactly match with any known universality classes, but they match with both 3D Heisenberg and 3D XY models. A large MCE was observed due to the persistent spin fluctuations over a broad temperature regime and can be attributed to magnetic frustration. The obtained values of critical exponents from the fitting of field dependent $\Delta S_{\rm m}^{\rm peak}$ and $RCP$ are very much in agreement with those resulting from critical analysis of magnetic isotherms near $T_{\rm C}$. The second-order nature of the phase transition was confirmed by Banerjee criteria and also from the variation of exponent $n(H,T)$ with temperature. Further, the low $T_{\rm C}$ with no thermal hysteresis, large isothermal entropy change, and huge $RCP$ make LMMO a promising candidate for magnetic refrigeration, especially for the liquefaction of hydrogen gas.

The temperature dependence of DC magnetization measured under ZFC and FC protocols showed a bifurcation around $20$~K, which indicates the onset of a glassy transition. This glass transition is further confirmed by the AC susceptibility measurement. The obtained fitting parameters from the relative shift in $T_{\rm f}$ using different theoretical models point towards the formation of cluster-glass in LMMO. Magnetic memory measured in both ZFC and FC processes shows significant memory effect, further supporting the cluster-class nature. We observed that the negative-$T$ cycle preserves the memory under temporary cooling, whereas in a positive-$T$ cycle, a small heating re-initializes the relaxation process. This asymmetric behavior in the relaxation process is explained by the hierarchical model and is another endorsement of the cluster glass behaviour.

\section{acknowledgments}
For financial support, we would like to acknowledge SERB, India bearing sanction Grant No.~CRG/2022/000997.


%
	
\end{document}